\documentclass[a4paper,11pt]{article}

\pdfoutput=1 
\usepackage[english]{babel}
\usepackage{microtype,xspace}
\usepackage[T2A]{fontenc}
\usepackage[utf8]{inputenc}	

\usepackage{jheppub} 
\usepackage{bm,amssymb,amsmath,amsfonts,slashed,graphicx,multirow,soul,mathtools,xspace,array,comment,color}
\hypersetup{colorlinks=true, allcolors=blue}  
\usepackage{adjustbox}
\usepackage{siunitx}
\usepackage{etoolbox}
\usepackage{bm} 
\usepackage{float}
\allowdisplaybreaks
\usepackage{ bbold }
\usepackage{subfigure}
\usepackage[normalem]{ulem} 
\usepackage{enumerate}
\usepackage{booktabs}
\usepackage{xparse}
\usepackage{etoolbox}
\usepackage{physics}
\usepackage{hypcap}
\usepackage{siunitx}
\usepackage{xcolor}
\usepackage{comment}
\usepackage[capitalise]{cleveref}

\usepackage{colortbl,arydshln} 
\newcommand{\nuMuC}{\nu\text{MuC}}
\newcommand{\ds}[1]{$\displaystyle{#1}$}
\def\be{\begin{equation}}
\def\ee{\end{equation}}

\newcommand{\GeV}{\text{GeV}}

\definecolor{defgolden}{rgb}{0.95,0.7125,0.}
\definecolor{deflightblue}{rgb}{0.363898,0.618501,0.782349}
\definecolor{defgreen}{rgb}{0,0.6,0}
\definecolor{defpurple}{rgb}{0.5,0,0.5}
\definecolor{defcyan}{rgb}{0.,0.5,0.5}
\definecolor{defbrown}{rgb}{0.6,0.3,0.1}
\definecolor{defdarkgray}{rgb}{0.35,0.35,0.35}
\definecolor{deflightgray}{rgb}{0.65,0.65,0.65}

\arxivnumber{} 

\title{\boldmath Quark mixing from muon collider neutrinos}

\author[1]{David Marzocca,}
\emailAdd{david.marzocca@ts.infn.it}

\author[2]{Francesco Montagno,}
\emailAdd{fmontagno@ifae.es}

\author[1]{Manuel Morales-Alvarado,}
\emailAdd{manuel.morales.alvarado@ts.infn.it}

\author[2,3]{\mbox{Andrea Wulzer}}
\emailAdd{awulzer@ifae.es}

\affiliation[1]{INFN, Sezione di Trieste, Via Bonomea 265, 34136, Trieste, Italy}

\affiliation[2]{Institut de F\'{\i}sica d'Altes Energies (IFAE), The Barcelona Institute of Science and Technology (BIST),
Campus UAB, 08193 Bellaterra, Barcelona, Spain}
\affiliation[3]{ICREA, Instituci\'o Catalana de Recerca i Estudis Avan\c{c}ats, 
Passeig de Llu\'{\i}s Companys 23, 
08010 Barcelona, Spain}

\abstract{
A high energy muon collider naturally produces a collimated beam of neutrinos for a fixed-target experiment at a dedicated far-forward facility. The high intensity and energy of the beam makes it ideally suited for astonishingly precise measurements of neutrino scattering on nucleons in the deeply inelastic regime, enabling the determination of the Cabibbo--Kobayashi--Maskawa~(CKM) quark mixing matrix. We assess the floor to the attainable sensitivity set by irreducible sources of uncertainties from the imperfect knowledge of the parton distribution (PDF) and fragmentation functions, showing that a strong improvement is possible well above current standards. As a by-product, our analysis also outlines extraordinary perspectives for a combined determination of the PDF. The results demonstrate the potential of a parasitic neutrino experiment at the muon collider, motivating detailed future studies. 
}

\begin{document}

\maketitle
\flushbottom

\section{Introduction}

The 2020 particle physics European Strategy update process triggered a new enthusiasm on muon colliders~\cite{Delahaye:2019omf} and led to the formation of the International Muon Collider Collaboration (IMCC) pursuing an R{\&}D programme towards a muon collider (MuC) with 10\,TeV energy in the centre of mass and high luminosity~\cite{Accettura:2023ked,InternationalMuonCollider:2024jyv,InternationalMuonCollider:2025sys}. The Snowmass~\cite{Narain:2022qud}, P5~\cite{P5:2023wyd} and NAS~\cite{NAS} recommendations outlined the ambition of building a muon collider in the USA and planned investments on the R{\&}D program in the coming decade. 

A vast literature including review articles~\cite{AlAli:2021let,Black:2022cth,Accettura:2023ked,InternationalMuonCollider:2024jyv,InternationalMuonCollider:2025sys} exists on the physics opportunities offered by the study of $\mu^+\mu^-$ collisions. Here we consider instead the possibility of exploiting the high-energy neutrinos produced by the muon beam decay for a fixed-target scattering experiment at a dedicated facility. Such a muon collider neutrino ($\nuMuC$) experiment would run independently and in parallel with the muon collider operation and with the data taking at the main detector.

The 5\,TeV muon beam decay produces few-TeV energy muon- and electron-type neutrinos with a relatively large probability to interact with matter, due to the energy enhancement of the neutrino cross sections. The dominant neutrino interaction in the target is Deep Inelastic Scattering (DIS) with the nucleons, at high (typically 20\,GeV) transferred momentum. This enables DIS studies in a kinematic regime where accurate theoretical predictions are possible within perturbative QCD, offering a sharp connection between the measurements and the underlying microscopic theory parameters. In the present paper we study how to exploit this connection for a determination of the CKM matrix elements that set the strength of charged-current neutrino-quarks interactions. On top of inclusive DIS, the determination also requires measurements of Semi-Inclusive (SIDIS) processes where bottom, charm or strange-flavoured hadrons are tagged in the final state. The SIDIS measurements are a form of flavour tagging that disentangles the contribution from the different quark flavour transitions. The set of DIS and SIDIS measurements under consideration enables the simultaneous determination of all the CKM elements barring those involving the top quark, which is not kinematically accessible.

CKM matrix elements determination provides a fundamental test of the Standard Model mechanism for the generation of the quark masses---through the CKM unitarity relations---and it is sensitive to new physics. Moreover the CKM elements are inputs for the Standard Model predictions of rare flavour-changing processes and their current precision is often a limiting factor for the sensitivity to new physics of these observables. Current CKM knowledge comes from low-energy measurements such as superallowed nuclear $\beta$ decays, hadronic  $\tau$ decays, and semileptonic decays of pions, kaons, $D$ and $B$ mesons. The 2024 world averages~\cite{ParticleDataGroup:2024cfk} for the six CKM elements of interest to our study are reported in Table~\ref{tab:CKMPDG}. Low-energy measurements have limited future prospects for improvement because they are close to the floor of systematic uncertainties. Our results, reported on the table, show that a future $\nuMuC$ experiment has instead  the potential to improve the CKM measurements precision above current standards. Furthermore, the $\nuMuC$ CKM determination from neutrino scattering is methodologically independent from the traditional low-energy approach.

\begin{table}[t]
    \centering
    \begin{tabular}{|c|c|c|c|c|c|c|}
    \hline
     \multirow{2}{*}{PoI} & \multirow{2}{*}{PDG} & \multicolumn{3}{c|}{Relative Precision [\%]} & \multicolumn{2}{c|}{Gain: PDG/${\nuMuC}$} \\
     \cline{3-7}
&   & PDG & ${\nuMuC}_3$ & ${\nuMuC}_{10}$ & ~3\,TeV~ & 10\,TeV \\
     \hline
     $|V_{ud}|$ & $0.97373 \pm 0.00031$ & 0.032 & 0.0036 & 0.0032 & 9 &10 \\ \hdashline
     $|V_{us}|$ & $0.22431 \pm 0.00085$ & 0.38 & 0.085 & 0.067 &4.5 &5.6 \\ \hdashline
     $|V_{cd}|$ & $0.221 \pm 0.004$ & 1.8 & 0.050 & 0.041 &36 &44 \\ \hdashline
     $|V_{cs}|$ & $0.975 \pm 0.006$ & 0.62 & 0.031 & 0.016 & 20 &39 \\ \hdashline
     $|V_{cb}|$ & $(41.1 \pm 1.2) \times 10^{-3}$ & 2.9 & 0.44 & 0.24 & 6.6 &12 \\ \hdashline
     $|V_{ub}|$ & $(3.82 \pm 0.20) \times 10^{-3}$ & 5.2 & 1.9 & 1.6 &2.7 &3.3 \\ 
     \hline
     $s_W^2$ & $0.23129 \pm 0.00004$ & 0.017 & 0.016 & 0.015 & 1.1&1.1 \\ 
     \hline
    \end{tabular}
    \caption{PDG averages \cite{ParticleDataGroup:2024cfk} for the Parameters of Interest (PoI) of our fit
    ---i.e., the absolute value of six CKM elements and the Weinberg angle---
    and the corresponding relative precision. 
    The $\nuMuC_3$ and $\nuMuC_{10}$ columns report the potential sensitivity of the $\nuMuC$ experiment at a 3 and a 10\,TeV MuC.
    The last columns shows the precision gain, i.e.\ the ratio of PDG and $\nuMuC$ uncertainties.
    }
    \label{tab:CKMPDG}
\end{table}

\begin{figure}[t]
\centering
    \includegraphics[width=0.6\textwidth]{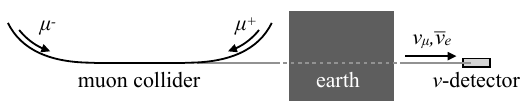} 
    \caption{Schematic of the $\nuMuC$ experimental setup.}
    \label{fig:expsetup}
\end{figure}

We consider a setup where the neutrinos from the decay of one of the beams in a straight section close to the interaction point are sent to a target surrounded by a detector, like in Figure~\ref{fig:expsetup}. The neutrinos are emitted with a very small angle from the decaying muons because of the large Lorentz boost. Muon decays in the straight section thus produce a collimated neutrino beam that can be collected efficiently by a compact target placed far enough from the collider. A first characterisation of the neutrino beam and a sketch of the experiment was presented in~\cite{InternationalMuonCollider:2024jyv,InternationalMuonCollider:2025sys} and it is revisited in Section~\ref{ssec:ExS}. Based on realistic collider parameters and under conservative assumptions, a neutrino experiment at the high-energy muon collider is found to deliver an unprecedented rate of neutrino-nucleon interactions that surpasses past experiments and future proposals. This enables neutrino DIS and SIDIS measurements with unprecedented statistics and at high energy.

TeV-energy neutrinos are produced also at the LHC from the decay of hadrons, and they are exploited for current neutrino experiments such as FASER~\cite{FASER:2019dxq,FASER:2022hcn} and SND@LHC~\cite{SHiP:2020sos,SNDLHC:2022ihg}, with proposed upgrades for the HL-LHC at the Forward Physics Facility (FPF)~\cite{Anchordoqui:2021ghd,Feng:2022inv}. These experiments have a much inferior DIS measurement statistics than $\nuMuC$. Furthermore, the energy spectrum and the composition of the muon collider neutrinos can be predicted with great accuracy, while neutrino production at hadron colliders is difficult to model resulting in large uncertainties. An accurately characterized, high-energy and high-intensity neutrino beam---ideally suited for precise DIS measurements---can be produced only at a high-energy muon collider. This offers an appealing long-term perspective for a radical improvement of the ongoing neutrino far-forward physics program, which is an active area of research (see e.g.~\cite{FPF:2025bor,John:2025qlm,FPFWorkingGroups:2025rsc,Adhikary:2024nlv,Kling:2023tgr,Ariga:2025qup,MammenAbraham:2024gun}).

The large statistics for DIS and SIDIS processes enables the measurement of the CKM with very small statistical uncertainties. The impact of parametric, theoretical and experimental uncertainties needs to be estimated for realistic sensitivity projections. In this paper, we start from the parametric uncertainties that are dominantly due to the imperfect knowledge of the Parton Distribution Functions~(PDF) and of the Fragmentation Functions~(FF), needed for the prediction of the DIS and SIDIS cross sections. These uncertainties provide an irreducible floor to the attainable sensitivity because they cannot be eliminated by, for instance, a more accurate calculation or a more performant detector. They could be only reduced by an independent determination of the PDF and the FF, which however cannot be envisaged at any present or future facility at the required level of accuracy. In fact, the best future prospects for PDF and FF determination are offered by the same $\nuMuC$ DIS and SIDIS datasets that we consider here for the CKM measurement~\cite{InternationalMuonCollider:2024jyv,InternationalMuonCollider:2025sys}. Our results emerge from a combined fit to the CKM and to the parameters of the PDF and of the FF, treated as nuisance parameters. 

Neutrino experiments at muon colliders or with dedicated muon rings of different energies were studied long ago\,\footnote{See also\,\cite{deGouvea:2025zfq} for a recent assessment of the electroweak physics potential of neutrino-electron scattering using the high-energy muon collider neutrino beam, and~\cite{Ball:2009mk} for a simultaneous determination of the CKM and the strange-quark PDF using NuTeV data.}~\cite{King:1997dx,Bigi:2001xb,Mangano:2001mj,Forte:2001zu}, leading in particular to a concept for the neutrino detector and target. Attempts were made to assess the physics potential, and in particular to guess the expected CKM measurement precision~\cite{Bigi:2001xb}. At variance with these studies, our projections are based on a quantitative analysis exploiting all the available experimental information for a CKM and PDF/FF combined fit. Furthermore, the usage of FF rather than the inherently qualitative notion of flavour tagging based on quark-hadron duality enables us to parametrize and to reduce, by the combined fit, the associated uncertainties.

The rest of this paper is organised as follows. Section~\ref{sec:MuCDIS} discusses the basic DIS and SIDIS phenomenology at the $\nuMuC$ experiment. This includes an estimate of the luminosity and the characterisation of the neutrino beam energy spectrum. We next describe the DIS and SIDIS cross section formulas---reported in Appendix~\ref{app:DISAPP}---and the observables that we employ in the analysis. We finally discuss---see also Appendix~\ref{app:pdf_ff}---the uncertainties on their prediction from the imperfect knowledge of the PDF and of the FF, which we model as nuisance parameters in a likelihood fit. In Section~\ref{sec:CKMd} we report our results for the $\nuMuC$ experiment potential. Section~\ref{subsec:FitRes} describes our CKM sensitivity projections as they emerge in our baseline analysis setup, while in Section~\ref{subsec:Fit-Anatomy} we assess the robustness of the results under analysis variants. A particularly important aspect is the dependence on the number of members in the PDF set, which reflects the flexibility of the PDF interpolation model. In Section~\ref{sec:PDFdet} we estimate the $\nuMuC$ experiment potential for the determination of the PDF, which emerges in our analysis from the determination of the nuisance parameters in the fit. The $\nuMuC$ experiments at the 10\,TeV muon collider is our main focus, however in Section~\ref{sec:3TeVres} we extend our analysis to a possible first stage of the muon collider with 3\,TeV energy in the centre of mass, obtaining very competitive results as in Table~\ref{tab:CKMPDG}. Conclusions and an outlook to future work are reported in Section~\ref{sec:conc}.

\section{Muon collider neutrino DIS}\label{sec:MuCDIS}

This section describes the phenomenology of deep inelastic scattering processes at the $\nuMuC$ experiment with neutrinos from the beam decay at a 10\,TeV MuC. In Section~\ref{ssec:ExS} we introduce a possible setup~\cite{InternationalMuonCollider:2025sys,InternationalMuonCollider:2024jyv}, aiming at an estimate of the `luminosity' (neutrino flux times the number of scattering centres) that can be envisaged for a $\nuMuC$ experiment parasitic to the muon collider. In Section~\ref{ssec:nuspectrum} we describe the simple and accurate calculation of the neutrino energy spectrum and in Section~\ref{ssec:NuDIS} we discuss the processes of interest for our analysis. These are the inclusive DIS and the SIDIS processes with bottom-, charm- or strange-flavoured hadrons in the final state. In Section~\ref{subsec:OBS} we introduce the observables that we employ in the fit and in Section~\ref{subsec:Parametric_Uncertainties} we describe the parametric uncertainties on their prediction from PDF and FF uncertainties. 

\subsection{Experimental setup}\label{ssec:ExS}

The muons that decay in a straight section of the collider ring produce a collimated beam of energetic neutrinos for a fixed-target experiment at a dedicated forward detector, as depicted in Figure~\ref{fig:expsetup}. The collider ring design foresees very short straight sections, except for the region close to the $\mu^+$ $\mu^-$ Interaction Point (IP) where a longer straight section is needed for the installation of the final focusing system and of the main muon collider detector. A 500\,m long straight section is foreseen in the current collider ring design~\cite{InternationalMuonCollider:2025sys,calzolari_2024_14000854}. However, a small portion of the straight section of 10~meter length is assumed in the following estimate for reasons that will become momentarily clear. The fraction of muons that decay in this region is equal to the length divided by the collider circumference of 10\,km\,\footnote{We consider Scenario~1 of Table~1.1.1 in Ref.~\cite{InternationalMuonCollider:2025sys}.}.

The collider operates with a single bunch of about $2\cdot10^{12}$ muons (plus one of anti-muons) that is injected every $0.2$ seconds and is left to circulate until when almost all the muons have decayed. This gives $10^{10}$ muon decays per second in the 10-meters straight region of interest, or $10^{17}$ decays in one year of muon collider operation, estimating one year of operation as \mbox{$10^7$s}. Each decay produces one usable neutrino of the muon type and one of the electron type. In what follow we consider the $\mu^-$ beam and hence muon-type neutrinos, $\nu_\mu$, and electron-type anti-neutrinos, $\bar\nu_e$. The $\mu^+$ beam could also be exploited, but this would require the installation of a second detector as the neutrinos from the two beams point in opposite directions. 

The typical energy of the neutrinos from the decay of a 5\,TeV muon is of order few~TeV, while their momentum transverse to the muon direction is less than half of the muon mass. The angular neutrino separation from the decaying muon is thus as small as $\Delta\theta\sim0.03$\,mrad. If the muons were parallel to the beam axis, the neutrino beam would be extremely collimated, with an angular spread $\langle\theta\rangle\sim0.03$\,mrad. Such small angle would be ideally suited for the experiment, as the majority of the neutrinos could be intercepted by a order 10\,cm radius cylindrical target placed 2\,km away from the IP and safely far from the collider ring. 

However, the muons are not parallel to the beam axis and the muon beam angular spread varies considerably along the straight section. The spread is as large as $0.8$\,mrad in the region close to the IP, where a large spread is needed in order to maximize the collider luminosity. The neutrinos produced in that region have an angular spread that is equal to the spread of the muon beam,  $\langle\theta\rangle\sim0.8$\,mrad. The optimal placement of the 10\,cm radius target for these neutrinos would be unrealistically close (about 100\,m) to the IP. If the target radius cannot be increased significantly (e.g., to 1\,m), not all the neutrinos produced in the $500$\,m long straight section will thus be useful for the experiment, but only those produced in the portion of the straight section where the beam angular spread is small. In our conservative estimate this portion is assumed to be only 10\,m long. The estimate will be soon superseded by a detailed calculation of the neutrino flux based on muon collider beam simulations~\cite{Calzolari2025Neutrino,Calzolari2025MDI}.

The rate for neutrino-nucleon collisions in the target is proportional to the neutrino flux times the number of scattering centres, which is the `luminosity' $\mathcal{L}$ of our fixed-target experiment. This quantity matches the more familiar collider definition of luminosity for two bunches with $N_{1,2}$ particles of species $1,2$ colliding at frequency $f_{\textrm{coll}}$ and with a cross-sectional area $\mathcal{A}$ of intersection: $\mathcal{L}=N_1 N_2 f_{\textrm{coll}}/\mathcal{A}=N_1\Phi_2$. The luminosity for neutrino-nucleon scattering is
\begin{equation}
    \mathcal{L}=N_N \Phi_\nu =\frac{R_\nu}{\mathcal{A}}\frac{\rho\,L\,{\mathcal{A}}}{M}=6\cdot 10^{33} {\textrm{cm}}^{-2}{\textrm{s}}^{-1}\frac{\rho\,L}{{\textrm{g}}/{\textrm{cm}}^2}
    =60\,\frac{{\textrm{fb}}^{-1}}{\textrm{year}}\frac{\rho\,L}{{\textrm{g}}/{\textrm{cm}}^2}\,,
\end{equation}
where $\rho$ is the mass density, $L$ is the length and ${\mathcal{A}}$---which however drops from the equation---is the area of the cylindrical target. The neutrino rate $R_\nu$ is estimated as $10^{10}/{\rm{s}}$ both for $\nu_\mu$ and ${\bar\nu}_e$ as previously explained. The nucleon mass is denoted as $M$.

The concept of a neutrino detector was described long ago in Ref.~\cite{King:1997dx}, for the neutrinos produced by a 250\,GeV muon ring. The target is a 1\,m long cylinder of 10\,cm radius and it consists of a stack of silicon CCD tracking planes. The arrangement of the planes corresponds to a mass per unit area $\rho L=50\,{\textrm{g}}/{\textrm{cm}}^2$, which gives instantaneous and integrated luminosities
\begin{equation}\label{eq:LowL}    \mathcal{L}_{\textrm{low}}=3\cdot10^{35}{\textrm{cm}}^{-2}{\textrm{s}}^{-1}\,,\qquad\int \hspace{-3pt}\mathcal{L}_{\textrm{low}}=15\,{\textrm{ab}}^{-1}.
\end{equation}
An operation time of 5 years is considered in the integrated luminosity matching the (conservative) baseline operation time of the muon collider. With a rough estimate of $10^{-35}{\textrm{cm}}^2=10\,{\textrm{pb}}$ for the cross section on nucleons of 1\,TeV energy neutrinos, this corresponds to 30~million recorded interactions per year and matches the `low' luminosity scenario---dubbed `10~Kg', because of the approximate target mass---described in Refs.~\cite{InternationalMuonCollider:2025sys,InternationalMuonCollider:2024jyv}.

A larger luminosity is found by considering more recent detector concepts proposed for the FPF program exploiting HL-LHC neutrinos~\cite{Anchordoqui:2021ghd,Feng:2022inv}. For instance, the FASER$\nu$2 detector target is made of tungsten and it has a length of $6.6$\,m. Given the tungsten density of $19.3\,{\textrm{g}}/{\textrm{cm}}^3$, this gives $\rho L=13\,{\textrm{Kg}}/{\textrm{cm}}^2$, which is 250 times more than the target in Ref.~\cite{King:1997dx}. However, it should be noted that the low density target is an integral part of the detector concept of Ref.~\cite{King:1997dx}. The neutrino scattering reactions occur inside the CCD silicon planes allowing the precise reconstruction of the event topologies from charged tracks and excellent capabilities for measurement and flavour tagging of the produced hadrons. The possibility of maintaining good experimental performances with a so much larger $\rho L$ should be investigated. On the other hand, the estimate in Eq.~(\ref{eq:LowL}) also relies on conservative assumptions on the expected rate of neutrinos and on the collider operation time. It thus appears realistic to envisage---like in~\cite{InternationalMuonCollider:2025sys,InternationalMuonCollider:2024jyv}---a `high' luminosity setup (dubbed `1~ton' in~\cite{InternationalMuonCollider:2025sys,InternationalMuonCollider:2024jyv}) with 
\begin{equation} \label{eq:HighL}
\int \hspace{-3pt}\mathcal{L}_{\textrm{high}}=100\int \hspace{-3pt}\mathcal{L}_{\textrm{low}}=1500\,{\textrm{ab}}^{-1}.
\end{equation}
From our study it emerges (see Figure~\ref{fig:luminosity-gainy}) that a strong motivation for targeting a high-luminosity $\nuMuC$ experiment is the perspective of a strong advance in the CKM determination.

\subsection{Neutrino energy spectrum}
\label{ssec:nuspectrum}

The probability distribution for the energy of the $\nu_\mu$ and of the $\bar\nu_e$ from the $\mu^-\to \nu_\mu \bar\nu_e e^-$ decay, at tree-level and neglecting the electron mass, reads\,\footnote{These formulas are obtained from the muon decay distribution $d\Gamma_\mu/dp_+$, with $p_+$ the plus component of the neutrino momentum, by identifying $p_+$ with the energy since the neutrinos are nearly parallel to the detector axis (the $z$ axis).} 
\begin{equation}\label{eq:epdf}
    \begin{split}
       {\mathcal{P}_{\nu_\mu}}(E)=\frac1{\Gamma_\mu}\frac{ d\Gamma_{\mu}}{ dE} =&       \frac1{3\bar{E}}\left[5-9 \left(\frac{E}{\bar{E}}\right)^2+4\left(\frac{E}{\bar{E}}\right)^3\right]\,,\\
       {\mathcal{P}_{\bar\nu_e}}(E)= \frac1{\Gamma_\mu} \frac{d\Gamma_{\mu}}{dE} =&     \frac2{\bar{E}}\left[1-3 \left(\frac{E}{\bar{E}}\right)^2+2\left(\frac{E}{\bar{E}}\right)^3\right]\,,
    \end{split}
\end{equation}
where $\bar{E}=(1+\beta)/2\, E_\mu$ is the maximal neutrino energy, denoting $\beta=(1-{m_\mu^2}/{E_\mu^2})^{1/2}$ the muon velocity. As the muon energy $E_\mu=5$\,TeV is very large, $\bar{E}$ is in practice equal to $E_\mu$.

The simple formulas in Eq.~(\ref{eq:epdf}) already provide an accurate description of the neutrino energy distribution, and the accuracy of the prediction could be easily improved. Theory uncertainties are below the percent level, and they can be lowered by orders of magnitude by including radiative corrections and the exact treatment of the electron mass. Notice that the finite angular acceptance of the target distorts the distribution significantly, but only at low energy because the neutrinos emitted at large angle have low energy. However, the distortion has a minor impact on the energy distribution of the scattering events, and can be safely ignored in the present study, since low-energy neutrinos have small cross section. This geometric effect and others associated with the dynamics of the beam in the entire straight section---and not only in the 10\,m long portion considered in our estimate---can be accurately modelled based on the muon beam simulation.

The perspectives for a precise knowledge of the muon collider neutrino beam should be contrasted with the large uncertainties on the energy flux and even on the flavour composition of the neutrinos produced in proton collisions. The uncertainties on the muon collider neutrino beam could play a role only if extremely accurate measurements will eventually turn out to be possible, and they can be safely neglected in the present exploratory study. One should  consider instead an uncertainty on the luminosity, which however does not play a role in our analysis that is based on the measurement of ratios of cross sections. 

\subsection{Neutrino DIS and SIDIS}\label{ssec:NuDIS}

The event rate for monochromatic $\nu_\mu$ or $\bar\nu_e$ with energy $E$ is the flux for the mono-energetic neutrinos, times the number of scattering centres times the cross section for the DIS or SIDIS process under consideration. By factoring out the total neutrino luminosity $\mathcal{L}$ we can thus define the differential cross section $d\sigma$ as
\begin{equation}\label{eq:Sigma}
    d\sigma = {\mathcal{P}_{\nu}}(E)\,d{E}\,d\hat\sigma\,,
\end{equation}
where $d\hat\sigma$ is the DIS or SIDIS differential cross section for fixed neutrino energy and ${\mathcal{P}_{\nu}}(E)$ is the neutrino energy probability distribution function in Eq.~(\ref{eq:epdf}). Notice that $d\sigma$ is differential in one more variable (the neutrino energy) than the regular DIS or SIDIS cross section for monochromatic neutrinos.

The cross sections depend on the PDFs of the different quarks and anti-quarks in the target nucleon. Our baseline sensitivity projections assume a `tungsten' nucleon, which approximately describes the scattering on a physical target made of tungsten like for instance in the FASER$\nu$2 experiment design. Ignoring nuclear effects and relying on the isospin symmetry, the corresponding PDFs are obtained from the PDFs of the proton as shown on the left panel of Table~\ref{eq:isoN}, setting $Z=74$ and $A=184$ that correspond to the atomic number and the average atomic mass of tungsten. We will also study---in Sections~\ref{subsec:Fit-Anatomy} and~\ref{sec:PDFdet}---the dependence on the nucleus composition by comparing with the results for pure proton or neutron target and for a silicon target ($Z=14$ and $A=28.1$). Silicon is almost but not exactly an isoscalar nucleon with $Z=A/2$, for which the up and down PDFs are equal. The exactly isoscalar target will be also considered for comparison. 

\begin{table}[t]
\centering
\begin{tabular}{|c|}
\hline
Effective nucleon PDFs\\
\hline
$\overset{{\scriptscriptstyle{(}}-{\scriptscriptstyle{)}}}{u}=\frac{Z}{A}\overset{{\scriptscriptstyle{(}}-{\scriptscriptstyle{)}}}{u}_p+\frac{A-Z}{A}\overset{{\scriptscriptstyle{(}}-{\scriptscriptstyle{)}}}{d}_{\hspace{-2pt}p}$\\[5pt]
$\overset{{\scriptscriptstyle{(}}-{\scriptscriptstyle{)}}}{d}=\frac{Z}{A}\overset{{\scriptscriptstyle{(}}-{\scriptscriptstyle{)}}}{d}_p+\frac{A-Z}{A}\overset{{\scriptscriptstyle{(}}-{\scriptscriptstyle{)}}}{u}_{\hspace{-2pt}p}$\\[5pt]
$\overset{{\scriptscriptstyle{(}}-{\scriptscriptstyle{)}}}{S} = \overset{{\scriptscriptstyle{(}}-{\scriptscriptstyle{)}}}{S}_{\hspace{-2pt}p}\,, \quad S = \{s,\, c,\, b\}$\\
\hline
\end{tabular}\hspace{50pt}
\begin{tabular}{|c|c|}
\hline
Nucleon & $A/Z$\\
\hline
\hline
Tungsten & $2.5$\\
\hline
Silicon & $2.01$\\
\hline
Isoscalar & $2$\\
\hline
Proton & $1$\\
\hline
Neutron & $\infty$\\
\hline
\end{tabular}
    \caption{{\bf{Left:}} The PDFs for a nucleon target with effective atomic number $Z$ and atomic mass $A$. {\bf{Right:}} The composition of the nucleons used in the paper. Tungsten is our baseline.} 
    \label{eq:isoN}
\end{table}

Several different neutrino/nucleon scattering processes are of interest for our analysis, as summarised at the end of the section in Table~\ref{tab:XSlist}. Charged-current (CC) processes are the production of a $\mu^-$ or of an $e^+$ for, respectively, the reactions initiated by a $\nu_\mu$ or a $\bar\nu_e$. Six types of processes are considered for each lepton flavour: inclusive DIS, where the flavour of the produced hadrons is not measured, and five SIDIS processes that require the presence in the final state of energetic hadrons with strange, charm or bottom flavour. The electric charge is resolved for the strange- and charm-flavoured hadrons while it is treated inclusively in the case of the bottom, resulting in five SIDIS categories. DIS and SIDIS neutral-current (NC) processes, not accompanied by the production of a charged lepton, are also considered. 

In addition to the neutrino energy $E$ in the lab frame, the processes of interest for our analysis are described---see Appendix~\ref{app:DISAPP}---by the regular DIS variables $x$ and $Q^2$, plus a fourth variable $z$ for the semi-inclusive processes. Note that $x$, $Q^2$ and $z$ can be measured in both CC and NC events because they only depend on the kinematic of the hadrons. Namely, they can be expressed as
\begin{equation}
\label{eq:kinvar0}
Q^2=2ME_h-W^2-M^2\,,\qquad
x=\frac{Q^2}{2M(E_h-M)}\,,\qquad z=\frac{E_{F}}{E_h-M}\,,
\end{equation}
where the `inelasticity' $W$ is the total invariant mass of the hadrons, $E_h$ is their total energy in the lab frame and $M$ is the nucleon mass. The energy of the strange, charm or bottom-flavoured hadron that is tagged in the SIDIS processes is denoted as $E_{F}$. The neutrino energy $E=E_\ell+E_h-M$ is instead observable only in the CC processes by measuring the energy $E_\ell$ of the charged lepton.

\begin{figure}[t]
    \centering
    \includegraphics[width=0.46\textwidth]{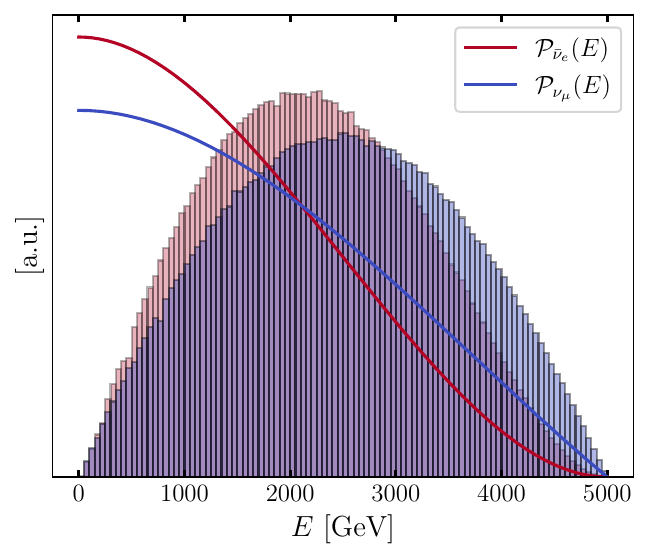}
    \hfill
    \includegraphics[width=0.5\textwidth]{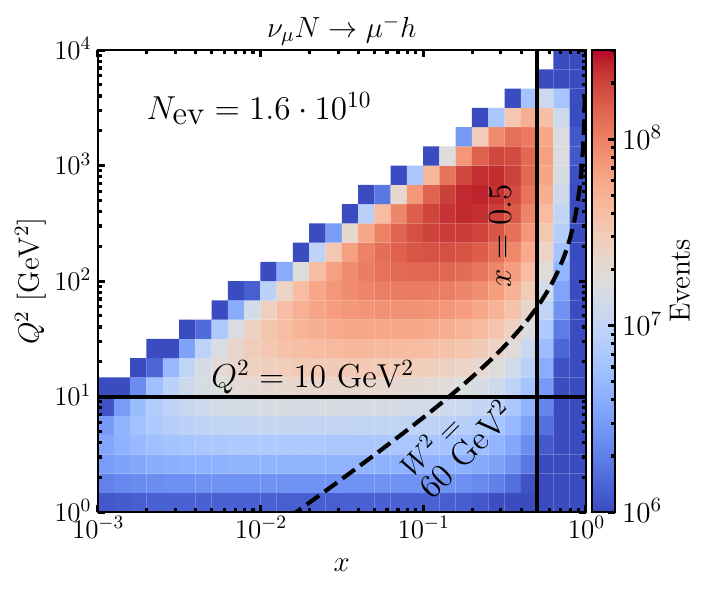}
    \caption{{\textbf{Left:}} Energy distribution of the interacting neutrinos, compared with the neutrino beam energy spectrum. {\textbf{Right:}} Number of events distribution (with high luminosity~(\ref{eq:HighL})) in $x$–$Q^2$ bins.}
    \label{fig:illustrative}
\end{figure}

\subsubsection{Inclusive DIS}
\label{ss:inc}

The regular (inclusive) DIS process is the production of hadrons associated with a charged lepton (in the CC case) or of an invisible neutrino (for NC). We assume a detector capable of distinguishing $\mu^-$ from $e^+$, therefore the CC DIS processes initiated by $\nu_\mu$ and by $\bar\nu_e$ are distinct experimentally. The corresponding triply-differential cross sections, defined as in Eq.~(\ref{eq:Sigma}), are reported in Eqs.~(\ref{eq:CCnu}) and~(\ref{eq:CCnubar}) at the tree-level order. Eq.~(\ref{eq:CCnu}) is repeated here for convenience:
\begin{equation}
\begin{aligned}
\frac{d\sigma({\nu_\mu N\to\mu^-h})}{dE dx dy} 
=   {\mathcal{P}}_{\nu_\mu} (E) \frac{2 G_F^2 M E x}{\pi(1+Q^2/m_W^2)^2} ~
&\bigg[\sum\limits_{\vphantom{\bar{d}}\textrm{I}=d,s,b}\sum\limits_{\vphantom{\bar{d}}\textrm{F}=u,c}
    |V_{\textrm{F}\textrm{I}}|^2 \textrm{I}(x,Q^2)+\\ 
    & \hspace{-40pt} +\sum\limits_{\vphantom{\bar{d}}\textrm{I}=\bar{u},\bar{c}}\sum\limits_{\vphantom{\bar{d}}\textrm{F}=\bar{d},\bar{s},\bar{b}}
    |V_{\textrm{I}\textrm{F}}|^2
    (1-y)^2
    {{\textrm{I}}}(x,Q^2)
     \bigg].
\end{aligned}
\end{equation}
The cross sections are expressed as double sums running over the initial and final state quarks or anti-quarks that are involved in the reaction, weighted by the relevant element of the CKM matrix `$V$' that controls the strength of the charged-current interactions. The top quark is not kinematically accessible and hence it does not appear in the final state. We work in the five-flavour scheme and therefore also the bottom is treated as a massless parton like the other quarks. 

Figure~\ref{fig:illustrative} shows the kinematic distributions for the CC scattering processes on an tungsten nucleon. The left panel is the distribution of the neutrino energy in the scattering events integrated in $x$ and $Q^2$, compared with the distribution~(\ref{eq:epdf}) of the neutrinos in the beam. While the beam spectrum extends to small energies, the energy of the interacting neutrinos is peaked at around 2.5\,TeV due to the linear growth with $E$ of the DIS cross section. The two-dimensional histogram on the right panel displays the binned events distribution in the $x$-$Q^2$ plane---integrated over the energy---of the $\nu_\mu$ scattering process. The total number events is about $10^{10}$ (see Table~\ref{tab:XSlist}) in the high luminosity setup with 1500\,ab$^{-1}$. The result is similar for the $\bar\nu_e$ process. The distribution is peaked at $x\sim0.2$ and $Q^2\sim400$\,GeV$^2$ and it is sharply localized---note that the colour bar is logarithmic---in a very favourable region of the plane. The inelasticity $W^2$ is typically above around 60\,GeV$^2$, deeply inside the DIS kinematic regime. Given that $Q^2$ is also large, perturbative QCD calculations enable accurate predictions that are directly linked to the microscopic theory parameters and in particular to the CKM matrix elements. 

The convergence of the perturbative expansion can be further improved by applying selection cuts. For our analysis we restrict to the region $Q^2>10\,{\textrm{GeV}}^2$ and we also impose $x<0.5$, with a twofold purpose. One is that the upper cut on $x$ enforces (given the $Q^2$ cut) a lower cut on $W^2$. The other is that the modelling of the PDFs from the sets that we employ in the analysis---see Section~\ref{subsec:Parametric_Uncertainties} and Appendix~\ref{app:pdf_ff}---is unreliable when $x$ is very close to 1. The cut on $x$ would not be needed with a better PDF modelling. At the real experiment, the finite detector resolution will produce contamination from the excluded region, however this effect is expectedly limited because the selection efficiency is very high. It is worth emphasising that such favourable conditions stem from the high energy of the neutrino beam (and in turn from the high muon beam energy) and cannot be attained by lower energy neutrino sources including low energy muon rings proposed in the past. Most of the paper focuses on the neutrino beam produced by the 10\,TeV muon collider, but the perspectives of a first stage of the muon collider operating at 3\,TeV are studied in Section~\ref{sec:3TeVres}.

The NC DIS final state is characterized by the production of hadrons not accompanied by a charged lepton. The flavour of the final-state neutrino cannot be determined, hence the cross section is the sum of the contributions from $\nu_\mu$ and $\bar\nu_e$ reported in Eqs.~(\ref{eq:NCnu}) and~(\ref{eq:NCnubar}). The only observables are $x$ and $Q^2$ since the neutrino energy cannot be measured. In Eqs.~(\ref{eq:NCnu}) and~(\ref{eq:NCnubar}), the sum over the initial state partons runs separately over quarks $q=u,c,d,s,b$ and anti-quarks $\bar{q}=\bar{u},\bar{c},\bar{d},\bar{s},\bar{b}$ and the final state parton is the same as the initial state one since the $Z$ interactions are diagonal. The $Z$ couplings, $g_f = T^3_L(f) - Q_f \sin^2 \theta_{\textsc{w}}$, depend on the Weinberg angle. The NC process is insensitive to the CKM, but it plays an important role in the CKM determination because it is sensitive to the PDFs. Furthermore, the NC measurements enable a determination of the Weinberg angle that improves current knowledge---see Table~\ref{tab:global_sensitivity_fit_unitaryCKM}---in a purely SM fit that exploits the unitarity of the CKM matrix. The total number of NC events---see later Table~\ref{tab:XSlist}---and their $x$-$Q^2$ distribution is similar to the CC ones.

\subsubsection{Semi-inclusive DIS}
\label{sec:SIDIS}

The semi-inclusive DIS (SIDIS) final states are characterized by the presence of a hadron $F$ with a given flavour that carries a significant fraction of the total hadrons energy. Specifically, $F$ is a $B$-hadron, a $D^*$ or a $K$ meson for bottom, charm or flavour-tagged SIDIS, respectively. The presence of the flavoured hadron is correlated with the flavour of the final-state quark in the hard scattering process. Therefore, the SIDIS measurements are effectively a (particularly simple) strategy for flavour tagging, which we need to perform in order to disentangle the different partonic transitions and in turn the different elements of the CKM matrix.

\begin{table}[t]
    \centering
    \begin{tabular}{|c||c||c||c||c|}
    \hline
     parton & bottom-tagged 
     & $c$-tagged 
     & $s$-tagged 
     & parton \\ \hline
      & $z\in [0.7,\,0.9]$ & $z\in [0.3,\,0.9]$ & $z\in [0.3,\,0.9]$ &  \\
     \hline \hline
     $b$ & \multirow{2}{*}{($5.63 \pm 0.07)\cdot 10^{-1}$} & \multirow{2}{*}{$(5.9 \pm 0.7)\cdot 10^{-2}$} & \multirow{2}{*}{$(1.44 \pm 0.08)\cdot 10^{-2}$} & $b$ \\ \cline{1-1} \cline{5-5} \hdashline
     $\bar b$ &  & & & $\bar b$ \\ \hline
     $c$ & \multirow{8}{*}{$(1.04 \pm 0.02)\cdot 10^{-4}$} & $(1.95 \pm 0.18) \cdot 10^{-1}$ & \multirow{2}{*}{$(6.21 \pm 0.10)\cdot 10^{-2}$} & $c$ \\ \cline{1-1} \cline{3-3} \cline{5-5} \hdashline
     $\bar c$ & & \multirow{7}{*}{$(5.8 \pm 1.5)\cdot 10^{-4}$} & & $\bar c$ \\ \cline{1-1} \cline{4-4} \cline{5-5} \hdashline
     $s$ & &  & $(1.80 \pm 0.09)\cdot 10^{-1}$ & $s$ \\ \cline{1-1} \cline{4-4} \cdashline{2-4} \cline{5-5}
     $\bar s$ & & & \multirow{2}{*}{$(6.2 \pm 1.4)\cdot 10^{-3}$} &  $\bar s$ \\ \cline{1-1} \hdashline
     \cline{5-5}
     $u$ & & & & $u$ \\ \cline{1-1} \cline{4-4} \cline{5-5}\cdashline{2-4} 
     $\bar u$ & & & $(4.44 \pm 0.18)\cdot 10^{-2}$ &  $\bar u$ \\ \cline{1-1} \cline{4-4} \cline{5-5} \hdashline
     $d$ & & & \multirow{2}{*}{$(1.28 \pm 0.34)\cdot 10^{-2}$} & $d$ \\ \cline{1-1} \cline{5-5} \hdashline
     $\bar d$ & & & & $\bar d$ \\ \hline
     $g$ & $(6.28 \pm 0.09)\cdot 10^{-3}$ & $(6.2 \pm 0.5)\cdot 10^{-3}$ & $(1.2 \pm 0.5)\cdot 10^{-2}$ & $g$ \\
     \hline
    \end{tabular}
    \caption{Efficiencies and mistag rates (with uncertainties) for each parton in different SIDIS processes. They are the integral of the corresponding FF integrated in the $z$ window reported in the second row, for $Q^2 = 400 \, \GeV^2$. Due to relations among the FFs, some values are identical for different partons and they are reported in a single row. The $\bar{c}$- and $\bar{s}$-tagged efficiencies and mistag rates can be obtained by charge conjugation.
    }
    \label{tab:TagEff}
\end{table}

The SIDIS cross sections are readily obtained by including the relevant Fragmentation Functions (FF) in the corresponding DIS cross section formulas, as in Eqs.~(\ref{eq:CCnuB}) and~(\ref{eq:CCnubarB}) for the CC and in Eqs.~(\ref{eq:NCnuB}) and~(\ref{eq:NCnubarB}) for the NC processes. The FFs are the final-state analogue of the PDFs and they are generically denoted as
\begin{equation}
    {\mathcal{F}}_{\textrm{F}}(z,Q^2)\,,
\end{equation}
where `${\textrm{F}}$' is a final-state parton (quark or gluon) and $z$ is approximately---see~(\ref{eq:kinvar0})---the fraction of the hadronic energy carried by the $F$ hadron. 

We restrict the SIDIS process final state to a $z$ window
\begin{equation}\label{eq:zwin}    z\in[z_{\min},\,z_{\max}]\,.
\end{equation}
Namely, scattering events where the $F$ hadron is inside the window are assigned to the SIDIS category with the flavour that corresponds to the one of the hadron, while the others are considered untagged. Different intervals are considered for the different flavour categories as reported on the second row of Table~\ref{tab:TagEff}. The upper bound of $z_{\max}=0.9$ is dictated by the availability of the FFs, while the lower bound is optimised for CKM parameters sensitivity.

In a tree-level calculation, the FFs can be interpreted as the probability to produce the $F$ hadron from a given parton in the final state. Therefore, the integral of the FFs in the $z$ window for each parton is the corresponding effective efficiency or mistag rate. The values are reported in Table~\ref{tab:TagEff} for the typical $Q^2 = 400 \, \GeV^2$ scale of the process. These efficiencies and mistag rates correspond to flavour tagging performances that are considerably inferior to those of more refined tagging strategies that are employed in future collider projections~\cite{Bedeschi:2022rnj,Blekman:2024wyf}. For instance, the light quarks and gluon mistag rates of the bottom tagger are one or two orders of magnitude larger than the ones of Refs.~\cite{Bedeschi:2022rnj,Blekman:2024wyf} for a comparable tagging efficiency of about $60\%$. Nevertheless, the conservative approach to flavour tagging based on SIDIS measurements is preferred because the SIDIS cross sections can be predicted in perturbative QCD relying on factorization. This offers good perspectives to obtain theoretical predictions that match the high accuracy needs of the analysis.

The different SIDIS flavour categories that are considered in the analysis are described in the rest of this section in turn. The modelling of the FFs and of their uncertainties, based on Refs.~\cite{Czakon:2022pyz,Anderle:2017cgl,AbdulKhalek:2022laj} is detailed in Appendix~\ref{app:pdf_ff}.

\subsubsection*{Bottom-tagged SIDIS}

\begin{figure}[t]
    \centering
    \includegraphics[width=0.49\textwidth]{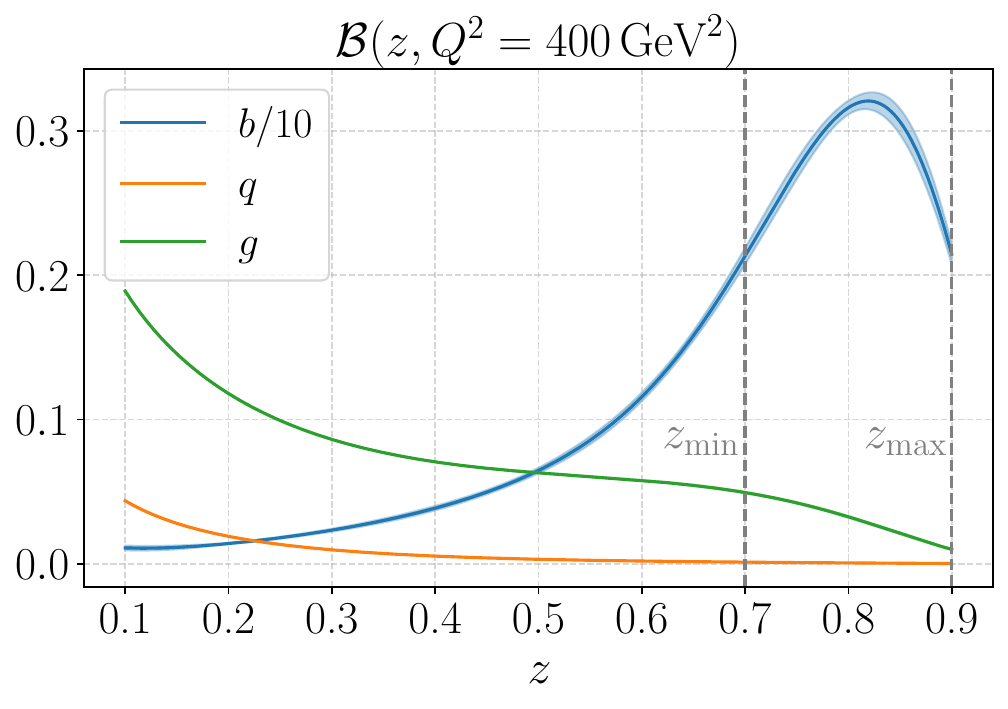} 
    \includegraphics[width=0.49\textwidth]{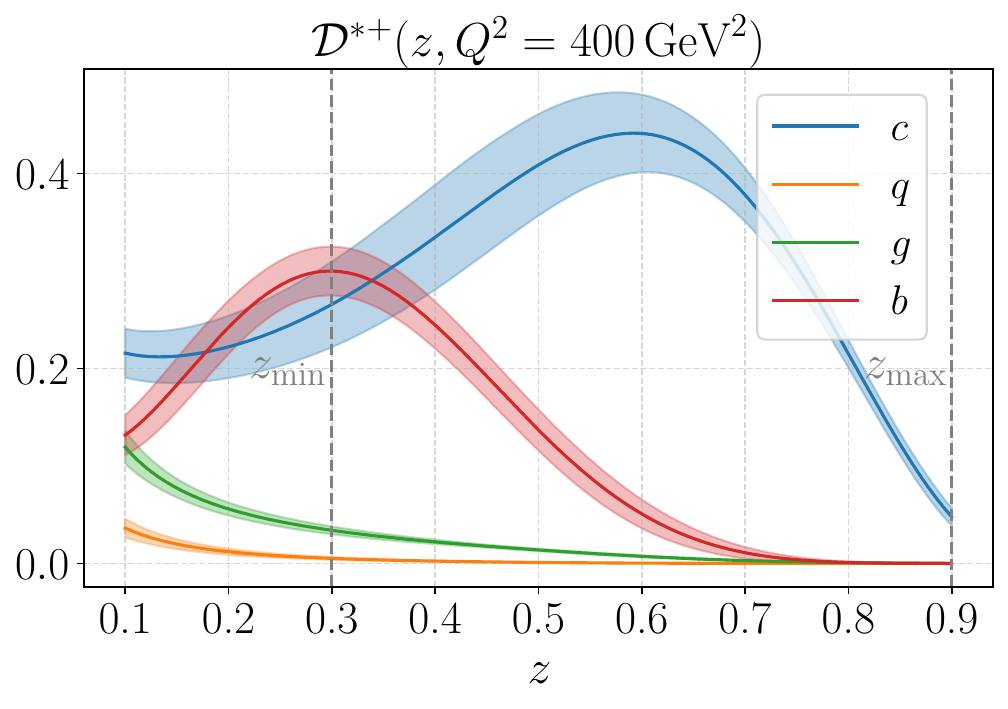}\\
    \includegraphics[width=0.50\textwidth]{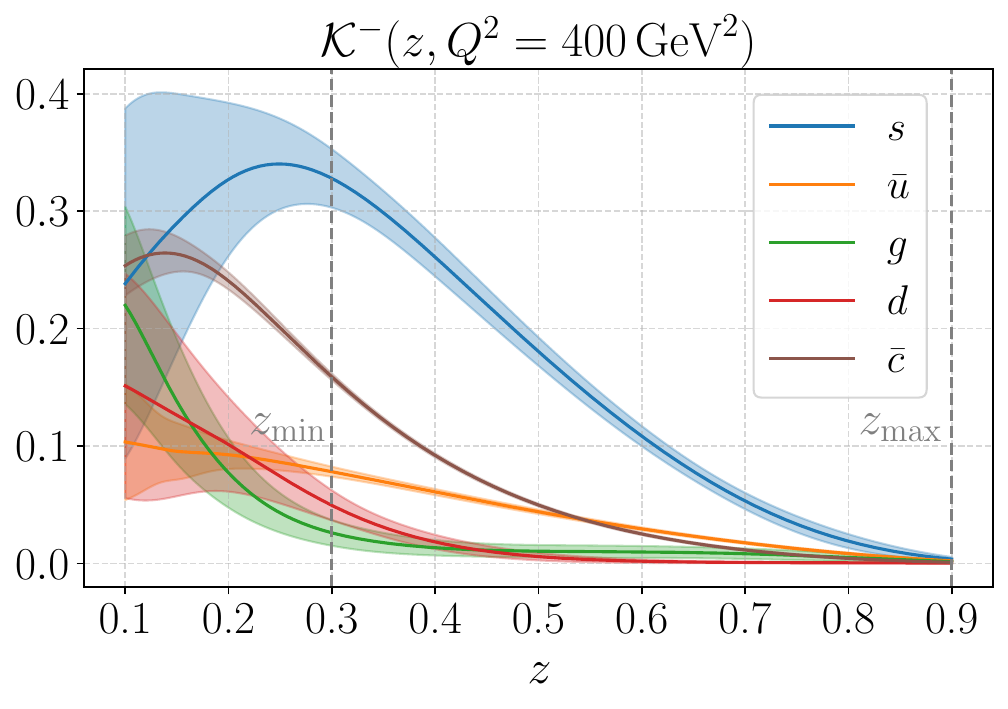}
    \caption{Fragmentation functions of $B$ (top left), $D^{*+}$ (top right) and $K^-$ (bottom) mesons for for different partons, with 68\% CL uncertainty bands. Quarks lighter than $b$ (top left) or all quarks except $b$, $\bar b$, and $c$ (top right) are indicated collectively as $q$. See the main text for details. The vertical dashed lines indicate the kinematic $z$ windows used in our analysis.}
    \label{fig:ffs}
\end{figure}

The relevant hadrons for tagging the bottom quark are the $B$-hadrons. The corresponding FFs~\cite{Czakon:2022pyz}, ${\mathcal{B}}_{\textrm{F}}(z,Q^2)$, are displayed in the top-left panel of Figure~\ref{fig:ffs} as a function of $z$ at $Q^2=400$~GeV$^2$. The figure reports the FF for the bottom quark, for any of the light quarks and for the gluon. In the framework of Ref.~\cite{Czakon:2022pyz}, the FF for all light quarks is the same and furthermore the PDF for all the anti-quarks (including the $\bar{b}$) is equal to the one of the corresponding quark because the charge of the $B$ hadron is not tagged. In turn, this means that the bottom-tagged SIDIS cross section measurement does not discriminate the $b$ from the $\bar{b}$ quark contribution. This is not a major limiting factor for our tree-level analysis, because only the $b$ or the $\bar{b}$ contributes, respectively, to the $e^+$ CC cross section~(\ref{eq:CCnubarB}) and to the $\mu^-$ CC cross section~(\ref{eq:CCnuB}).

Bottom-tagged SIDIS measurements are needed in order to access $u\to b$ and $c\to b$ partonic transitions (and their charge conjugate) in order to get sensitivity to $V_{ub}$ and $V_{cb}$. In inclusive DIS, the $u\to b$ and $c\to b$ contributions are very small---due to the smallness of $V_{ub}$ and $V_{cb}$---and the cross section is dominated by hard scattering processes with light quarks in the final state. An aggressive $z_{\min}=0.7$ cut on $z$ is thus needed for a strong enough suppression of the light quarks background. With this cut (and the upper limit $z_{\max}=0.9$, above which the FFs are not available) the light quarks mistag probability is about $0.01\%$ while the $b$-quark tagging efficiency is as high as around $60\%$. The total number of events is of about two millions in the high luminosity setup of Eq.~(\ref{eq:HighL}), as shown in Table~\ref{tab:XSlist}. The $c\to b$ and ${\bar{c}}\to {\bar{b}}$  transitions give an order-one contribution the bottom-tagged SIDIS cross sections while the contribution from $u\to b$ is at the few percent level (and the one of ${\bar{u}}\to {\bar{b}}$ is negligible). Therefore, a better relative precision is expected on $V_{cb}$ than on $V_{ub}$ compatibly with our final results.

The mistag rate for the gluon is $0.6\%$, much higher than the one for quarks. Therefore, even if gluon production only emerges at the NLO in the QCD expansion, it could give a significant contribution to the bottom-tagged SIDIS cross sections. This contribution is effectively a background for the determination $V_{ub}$ and $V_{cb}$ that is not included in our tree-level results. While the effect of this gluon background can be only assessed by an NLO analysis that is beyond the scope of the present paper, we notice that its impact could be reduced by charge-tagged SIDIS measurements that discriminate the $b$ from the $\bar{b}$ because the gluon contributes the same to the two charge categories.

\subsubsection*{{$\boldsymbol{c}$}-tagged and {${\boldsymbol\bar{\boldsymbol{c}}}$}-tagged SIDIS}

Two specific charm-flavoured hadrons are considered in the charm-tagged SIDIS final state~\cite{Anderle:2017cgl}: the $D^{*+}$ and the $D^{*-}$ meson. Two ($c$-tagged and $\bar{c}$-tagged) SIDIS categories are defined, depending on the charge of the $D^{*}$ meson. The corresponding FFs
\begin{equation}  {\mathcal{D}}^+_{\vphantom{{\overline{\textrm{F}}}}\textrm{F}}(z,Q^2)={\mathcal{D}}^-_{\overline{\textrm{F}}}(z,Q^2)\,,
    \label{eq:ff_cc}
\end{equation}
are related by charge conjugation. The $D^{*+}$ meson FFs, ${\mathcal{D}}^+_{\textrm{F}}(z,Q^2)$, are shown in the top-right panel of Figure~\ref{fig:ffs} for the charm quark, the bottom, the gluon and any other quark or anti-quark that are all assumed to have the same FF, including the $\bar{c}$. 

The charm-tagged SIDIS cross sections are useful to access $d\to{c}$ and $s\to{c}$ transitions and in turn $V_{cd}$ and $V_{cs}$. In inclusive DIS, these contributions are subdominant but not extremely small as $V_{cd}$ and $V_{cs}$ are sizeable. Therefore the charm-tagging selection window ($z_{\min} = 0.3$ and $z_{\max} = 0.9$) can be more inclusive than the bottom-tagging window because less background reduction is needed. With this window, the efficiency for the tagging of the $c$ quark is of $20\%$ and the mistag rate is about $0.05\%$ for light quarks, $6.5\%$ for the bottom and $0.63\%$ for the gluon. The probability for a $\bar{c}$ quark to produce a $D^{*+}$ meson is as small as for the light quarks, making the $D^{*+}$ meson final state selection an efficient tagger of positive-charge $c$-quarks. The $D^{*-}$ mesons selection tags instead the $\bar{c}$ anti-quark. The bottom quark mistag rate is only a factor of few smaller than the charm quark efficiency. However, the contribution to the charm-tagged CC cross section from $b$ quarks is suppressed by the smallness of $V_{cb}$ and $V_{ub}$ and completely negligible. 

Tagging the charge of the $D^*$ meson and the charge (and flavour) of the leptons enables the measurement of four CC SIDIS categories, listed in Table~\ref{tab:XSlist}. Only the sum of the $\nu_\mu$ and $\bar\nu_e$ processes is observable, leading to a total of two ($c$-tagged and $\bar{c}$-tagged) NC measurements. Order 100 million events are expected in the final states $\mu^-D^{*+}$ and $e^+D^{*-}$. These cross sections are dominated by $d\to c$ and $s\to c$ transitions and those with anti-quarks. The cross sections for $\mu^-D^{*-}$ and $e^+D^{*+}$ are much smaller because no $c$ ($\bar{c}$) quark can be produced in association with the $e^+$ ($\mu^-$) such that all the events in these categories emerge from mistagging. Since they are small, and furthermore they are not sensitive to $V_{cd}$ and $V_{cs}$, these measurements are excluded from the analysis. However, they could be useful in order to suppress the gluon background that emerges at NLO like in the case of the bottom.

\subsubsection*{{$\boldsymbol{s}$}-tagged and {${\boldsymbol\bar{\boldsymbol{s}}}$}-tagged SIDIS}

Strange tagging is based on the detection of a $K^-$ or a $K^+$ meson, which is used to define the $s$- and the $\bar{s}$-tagged SIDIS categories, respectively. The corresponding FFs~\cite{AbdulKhalek:2022laj} are related by charge conjugation
\begin{equation}  {\mathcal{K}}^+_{\vphantom{{\overline{\textrm{F}}}}\textrm{F}}(z,Q^2)={\mathcal{K}}^-_{\overline{\textrm{F}}}(z,Q^2)\,,
\end{equation}
and they are reported (for negative-charge kaon, corresponding to $s$-tagging) in the bottom panel of Figure~\ref{fig:ffs} for some of the partons. In the approach of Ref.~\cite{AbdulKhalek:2022laj}, 7 FFs are determined independently by imposing the constraints that the ${\mathcal{K}}^-$ FFs are equal (at the input $Q^2$ scale) for the following pairs of quark/antiquark: $(b,\bar{b})$, $(c,\bar{c})$, $(d,\bar{d})$ and $(u,\bar{s})$.

The FFs for the up and down quarks are only a factor few smaller than the one of the strange, hence the SIDIS process with kaons in the final state is not a high purity strange tagger. This is not an issue for our analysis because the flavour transitions with the strange in the final state (in particular, $u\to{s}$) are relatively sizeable already in the inclusive DIS process. A low-purity strange tagger is thus sufficient because all what is needed is to define observables that are sensitive to different combinations of $\{V_{ud},V_{us},V_{cd},V_{cs}\}$ than the ones entering in the DIS and in the charm-tagged SIDIS processes, resolving flat directions. An inclusive selection window of $z_{\min}=0.3$ to $z_{\max}=0.9$ is sufficient for this purpose. The corresponding efficiency and mistag rates are reported in Table~\ref{tab:TagEff}. In our analysis we employ all the four $s$- and $\bar{s}$-tagged CC processes with $e^+$ and $\mu^-$ in the final state and the two $s$- and $\bar{s}$-tagged NC processes. About 100 million events are expected in each category as shown in Table~\ref{tab:XSlist}.

\subsection{Observables}
\label{subsec:OBS}

We base our sensitivity projections on measurements of the cross section in kinematic bins for the 16 DIS and SIDIS processes listed in Table~\ref{tab:XSlist}. We label the processes according to their final state, such that for instance `$eh$' represents $e^+$ production in inclusive DIS and `$\mu D^{*+}$' is the $c$-tagged SIDIS $\mu^-$ production. The cross section of the NC processes such as $\nu h$, $\nu B$, etc.\,are the sum of the cross sections for $\nu_\mu$ and $\bar\nu_e$ scattering. 

\begin{table}[t]
    \centering
    \begin{tabular}{|c|c|c|c|c|c|c|}
\hline
 & \multicolumn{2}{c|}{\textbf{Inclusive}} & \multicolumn{2}{c|}{\textbf{$\boldsymbol{s}$-tagged}} & \multicolumn{2}{c|}{\textbf{$\boldsymbol{\bar s}$-tagged}} 
\\
\hline
 & Process & $N_{\rm{ev}}$ & Process & $N_{\rm{ev}}$ & Process & $N_{\rm{ev}}$ 
\\
\hline
\multirow{2}{*}{\textbf{CC}} & $\nu_\mu W\hspace{-2pt}\to\hspace{-2pt} \mu^-h $ & $1.5\hspace{-1pt}\cdot\hspace{-1pt}10^{10}$ & $\nu_\mu W\hspace{-2pt}\to\hspace{-2pt} \mu^-K^- $ & $2.4\hspace{-1pt}\cdot\hspace{-1pt}10^{8}$ & $\nu_\mu W\hspace{-2pt}\to\hspace{-2pt} \mu^-K^+ $ & $7.7\hspace{-1pt}\cdot\hspace{-1pt}10^{8}$  \\ 
\cline{2-7} 
& $\bar\nu_e W\hspace{-2pt}\to\hspace{-2pt} e^+h $ & $6.4\hspace{-1pt}\cdot\hspace{-1pt}10^{9}$ & $\bar\nu_e W\hspace{-2pt}\to\hspace{-2pt} e^+K^- $ & {$2.9\hspace{-1pt}\cdot\hspace{-1pt}10^{8}$} & $\bar\nu_e W\hspace{-2pt}\to\hspace{-2pt} e^+K^+ $ &  $1.5\hspace{-1pt}\cdot\hspace{-1pt}10^{8}$  \\
\hline
\multirow{2}{*}{\textbf{NC}} & $\nu_\mu W\hspace{-2pt}\to\hspace{-2pt} \nu_\mu h $ & \multirow{2}{*}{$7.5\hspace{-1pt}\cdot\hspace{-1pt}10^{9}$} & $\nu_\mu W\hspace{-2pt}\to\hspace{-2pt} \nu_\mu K^- $ & \multirow{2}{*}{$1.9\hspace{-1pt}\cdot\hspace{-1pt}10^{8}$} & $\nu_\mu W\hspace{-2pt}\to\hspace{-2pt} \nu_\mu K^+ $ &\multirow{2}{*}{$2.5\hspace{-1pt}\cdot\hspace{-1pt}10^{8}$}  \\ 
& $\bar\nu_e W\hspace{-2pt}\to\hspace{-2pt} \bar\nu_e h $ &  & $\bar\nu_e W\hspace{-2pt}\to\hspace{-2pt} \bar\nu_e K^- $ &  & $\bar\nu_e W\hspace{-2pt}\to\hspace{-2pt} \bar\nu_e K^+ $ &    \\
\hline
 & \multicolumn{2}{c|}{\textbf{$\boldsymbol{c}$-tagged}} & \multicolumn{2}{c|}{\textbf{$\boldsymbol{\bar c}$-tagged}} & \multicolumn{2}{c|}{\textbf{bottom-tagged}} 
\\
\hline
 & Process & $N_{\rm{ev}}$ & Process & $N_{\rm{ev}}$ & Process & $N_{\rm{ev}}$ 
\\
\hline
\multirow{2}{*}{\textbf{CC}} & $\nu_\mu N\hspace{-2pt}\to\hspace{-2pt} \mu^-D^{*+} $ & $4.4\hspace{-1pt}\cdot\hspace{-1pt}10^{8}$ & $\color{gray}\nu_\mu W\hspace{-2pt}\to\hspace{-2pt} \mu^-D^{*-} $ &\color{gray} $8.0\hspace{-1pt}\cdot\hspace{-1pt}10^{6}$ & $\nu_\mu W\hspace{-2pt}\to\hspace{-2pt} \mu^-B $ &$1.7\hspace{-1pt}\cdot\hspace{-1pt}10^{6}$  \\ 
\cline{2-7} 
& $\color{gray}\bar\nu_e W\hspace{-2pt}\to\hspace{-2pt} e^+D^{*+} $ &\color{gray} $3.3\hspace{-1pt}\cdot\hspace{-1pt}10^{6}$ & $\bar\nu_e W\hspace{-2pt}\to\hspace{-2pt} e^+D^{*-} $ & $2.6\hspace{-1pt}\cdot\hspace{-1pt}10^{8}$ & $\bar\nu_e W\hspace{-2pt}\to\hspace{-2pt} e^+B $ & $8.3\hspace{-1pt}\cdot\hspace{-1pt}10^{5}$   \\
\hline
\multirow{2}{*}{\textbf{NC}} & $\nu_\mu W\hspace{-2pt}\to\hspace{-2pt} \nu_\mu D^{*+} $ &\multirow{2}{*}{ $3.7\hspace{-1pt}\cdot\hspace{-1pt}10^{7}$} & $\nu_\mu W\hspace{-2pt}\to\hspace{-2pt} \nu_\mu D^{*-}$ & \multirow{2}{*}{$3.5\hspace{-1pt}\cdot\hspace{-1pt}10^{7}$} & $\nu_\mu W\hspace{-2pt}\to\hspace{-2pt} \nu_\mu B $ &\multirow{2}{*}{ $5.2\hspace{-1pt}\cdot\hspace{-1pt}10^{7}$}  \\ 
& $\bar\nu_e N\hspace{-2pt}\to\hspace{-2pt} \bar\nu_e D^{*+} $ &  & $\bar\nu_e W\hspace{-2pt}\to\hspace{-2pt} \bar\nu_e D^{*-} $ & & $\bar\nu_e W\hspace{-2pt}\to\hspace{-2pt} \bar\nu_e B$ &   \\
\hline
\end{tabular}
    \caption{The DIS and SIDIS processes with the corresponding total number of expected events in the high-luminosity setup~(\ref{eq:HighL}) and with a tungsten nucleon (denoted as $W$). Only the events passing the analysis cuts are included. The two processes in grey are not considered for the analysis.}
    \label{tab:XSlist}
\end{table}

The huge data statistics would enable an extremely fine binning, however a reasonable binning strategy should take into account the finite resolution in the experimental measurement of the kinematic variables. In particular, the bin size should be larger or comparable with the resolution in order to avoid a massive contamination in the measured-level bin counts from nearby truth-level bins, which would prevent unfolding. An experimental resolution better than about 10\,\% can be safely assumed, which corresponds to about 10 bins covering one order of magnitude variation of the variable. Specifically, we employ 30 logarithmically-spaced bins in $x$ from $10^{-3}$ to 1, and 24 logarithmic bins in $Q^2$ from $10\,\GeV^2$ to $10^4\, \GeV^2$, covering the relevant region---see Figure~\ref{fig:illustrative}---of the $x$-$Q^2$ plane. Our modelling of the PDF---see Section~\ref{subsec:Parametric_Uncertainties} and Appendix~\ref{app:pdf_ff}---is not reliable when $x$ is very close to~1. For this reason, only the first 27 bins in $x$ are included in the analysis, covering up to $x\simeq0.5$. 

A similarly fine binning could be considered for the variable $z$ in SIDIS processes. This would enable an accurate determination of the FFs and possibly also improve the determination of the parameters of interest. However, our modelling of the FFs---in particular, for the charm---is not sufficiently expressive for a credible determination: its few free parameters could likely be fitted with very good accuracy potentially producing optimistic results. For this reason, no binning is performed on the $z$ variable and the whole range $[z_{\min},\,z_{\max}]$ (as specified in Table~\ref{tab:TagEff}) is considered inclusively. Only bins in $x$ and $Q^2$ are employed for the NC processes and the corresponding cross sections are denoted (in Table~\ref{tab:XSlist}) as 2-index tensors $\sigma_{ij}$. In the CC processes the neutrino energy $E$ is also observable. Five linearly-spaced bins in $E$ are considered ranging from 0 to 5\,TeV. The CC cross sections are binned in 3 variables and are denoted as 3-index tensors $\sigma_{ijk}$. 

For our analysis we need the theoretical predictions for the binned cross sections $\sigma$ as a function of the parameters of interest of the fit. In the case of the CC processes, the parameters of interest are the CKM matrix elements\,\footnote{Notice that taking $G_F$, $m_Z$ and $\sin \theta_{\textsc{w}}$ as the SM input parameters, the CC cross sections are also sensitive to $\sin \theta_{\textsc{w}}$, through $m_W$. However, this dependence is suppressed by $Q^2/m_W^2$ and safely negligible. The present-day uncertainties on the remaining SM parameters $G_F$ and $m_Z$ are too small to play a role.} and the computation of $\sigma$ proceeds as follows. We compute separately, using \texttt{Vegas}~\cite{Lepage:1977sw} for the Monte Carlo integration, the coefficients of $|V_{\textrm{F}\textrm{I}}|^2$ and $|V_{\textrm{I}\textrm{F}}|^2$ in the DIS cross section formulas of Eqs.~(\ref{eq:CCnu}) and~(\ref{eq:CCnubar}). From these integrals we obtain the coefficients of the linear dependence of the DIS cross section on  $|V_{ij}|^2$. The same integrals can be employed for the determination of the SIDIS cross sections---see Eqs.~(\ref{eq:CCnuB}) and~(\ref{eq:CCnubarB})--- by noticing that the dependence on $Q^2$ of the FFs is weak so that the FFs are approximately constant in the narrow $Q^2$ bins that we employ. We can thus compute the integral in $z$ of each FF ${\mathcal{F}}_{\textrm{F}}(z,Q^2)$, obtaining a vector of efficiency/mistag factors (for each bin) to be multiplied by the corresponding DIS cross section integral coefficient and CKM element. The efficiencies and mistag rates shown in Table~\ref{tab:TagEff} are a particular case, evaluated at $Q^2 = 400 \,\GeV^2$. A similar approach is adopted for the calculation of the NC DIS and SIDIS cross sections in Eqs.~(\ref{eq:NCnu}), (\ref{eq:NCnubar}), (\ref{eq:NCnuB}) and~(\ref{eq:NCnubarB}) in order to reconstruct the dependence on the last parameter of interest, the sine of the Weinberg angle $\sin \theta_{\textsc{w}}$.

\begin{table}[t]
    \centering
\begin{tabular}{|c|cccc|c|}
\hline
& \multicolumn{4}{c|}{Observable}  &  Binning\\ 
\hline
\multirow{5}{*}{CC} & 
\multicolumn{4}{c|}{%
\rule{0pt}{5ex}\ds{R^{\mu/e}_{ijk} = \frac{\sigma^{\mu h}_{ijk}}{\sigma^{e h}_{ijk}}}%
} & 
\multirow{5}{*}{\parbox{60pt}{\centering $(x,\,Q^2,\,E)$ \\ $27\times24\times5$}} \\
& \multicolumn{4}{c|}{%
\ds{R^{eb}_{ijk} = \frac{\sigma^{e B}_{ijk}}{\sigma^{e h}_{ijk}}}\quad  
\ds{R^{\mu b}_{ijk} = \frac{\sigma^{\mu B}_{ijk}}{\sigma^{\mu h}_{ijk}}}\quad  
\ds{R^{e \bar{c}}_{ijk} = \frac{\sigma^{e D^{*-}}_{ijk}}{\sigma^{e h}_{ijk}}}\quad  
\ds{R^{\mu c}_{ijk} = \frac{\sigma^{\mu D^{*+}}_{ijk}}{\sigma^{\mu h}_{ijk}}}%
} & \\
& \multicolumn{4}{c|}{%
\ds{R^{es}_{ijk} = \frac{\sigma^{e K^-}_{ijk}}{\sigma^{e h}_{ijk}}}\quad  
\ds{R^{\mu s}_{ijk} = \frac{\sigma^{\mu K^-}_{ijk}}{\sigma^{\mu h}_{ijk}}}\quad  
\ds{R^{e \bar{s}}_{ijk} = \frac{\sigma^{e K^+}_{ijk}}{\sigma^{e h}_{ijk}}}\quad  
\ds{R^{\mu \bar{s}}_{ijk} = \frac{\sigma^{\mu K^+}_{ijk}}{\sigma^{\mu h}_{ijk}}}%
} & \\ [3ex]
\hline

\multirow{5}{*}{NC} & 
\multicolumn{4}{c|}{%
\rule{0pt}{5ex}\ds{R^{{\textsc{cc}}/{\textsc{nc}}}_{ijk} = \sum_k\frac{\sigma^{e h}_{ijk}+\sigma^{\mu h}_{ijk}}{\sigma^{\nu h}_{ij}}}%
} & \multirow{5}{*}{\parbox{60pt}{\centering $(x,\,Q^2)$ \\ $27\times24$}} \\
& \multicolumn{4}{c|}{%
\ds{R^{\nu b}_{ij} = \frac{\sigma^{\nu B}_{ij}}{\sigma^{\nu h}_{ij}}}\quad
\ds{R^{\nu c}_{ij} = \frac{\sigma^{\nu D^{*+}}_{ij}}{\sigma^{\nu h}_{ij}}}\quad
\ds{R^{\nu \bar{c}}_{ij} = \frac{\sigma^{\nu D^{*-}}_{ij}}{\sigma^{\nu h}_{ij}}}%
} &  \\
& \multicolumn{4}{c|}{%
\ds{R^{\nu s}_{ij} = \frac{\sigma^{\nu K^-}_{ij}}{\sigma^{\nu h}_{ij}}}\quad
\ds{R^{\nu \bar s}_{ij} = \frac{\sigma^{\nu K^+}_{ij}}{\sigma^{\nu h}_{ij}}}%
} & \\  [3ex]
\hline
\end{tabular}
    \caption{The observables employed in the fit.}
    \label{tab:Oblist}
\end{table}

The binned cross sections of the 16 DIS and SIDIS processes are not used directly in the fit. The analysis is on the contrary based on the 15 ratios of cross sections listed in Table~\ref{tab:Oblist}. This is because QCD corrections are expected to cancel to a large extent in the ratios, possibly enabling to reach the high theoretical accuracy that is required for the analysis. Specifically, the first CC observable is the ratio between muon and electron inclusive DIS. The other eight CC observables are formed as the bottom, charm, or strange flavour-tagged SIDIS cross section divided by the inclusive DIS cross section for electron or muon. The observables that depend on the NC cross sections are the `${{\textsc{cc}}/{\textsc{nc}}}$' ratio of the inclusive CC DIS cross section (the sum of electrons and muons) divided by the NC inclusive DIS cross section, and five NC SIDIS/DIS ratios.

The statistical errors on the observables are computed by error propagation, starting from the cross section measurement errors. The inclusive DIS cross section measurement features (small) statistical correlations with the SIDIS measurements in the same bin. The truly statistically independent variable is the `untagged' cross section, which is the DIS cross section minus the sum of the SIDIS cross sections in the different tagging categories. The number of expected events in each bin is in general very high, hence the measurements are effectively Gaussian distributed. However, the event yield in some category (typically, the bottom-tagged one) becomes small in some of the bins close to the kinematic boundaries and in this case the corresponding measurement is excluded from the analysis. Specifically, we exclude those observables whose measurement relies on less than 100 events.

The error on the statistically independent cross section measurements, performed by event counting, is the square root of the cross section prediction evaluated at the central value (from the PDG) of the parameters of interest and of the nuisance parameters, divided by the square root of the integrated luminosity.
The typical relative uncertainties in each bin are of the order of $10^{-4}$ for the inclusive DIS ratios $R^{\mu/e}$ and $R^{{\textsc{cc}}/{\textsc{nc}}}$, $10^{-3}$ for the charm and strange tagging ratios and for the NC $R^{\nu b}$ ratio. They are about $10^{-2}$ for the CC $b$-tagging ratios due to the smalness of the bottom-tagged cross sections.

\subsection{Parametric uncertainties}
\label{subsec:Parametric_Uncertainties}

We aim at performing a Gaussian likelihood fit with the observables listed in Table~\ref{tab:Oblist}, which we collect in a high-dimensional vector ${\mathcal{O}}$ flattening over the bins. The parameters of interest of the fit, denoted as `$\pi$' in what follows, are the six CKM matrix elements and the sine of the Weinberg angle. The observables are functions, ${\mathcal{O}}={\mathcal{O}}[\sigma]$, of the binned cross sections $\sigma$. The theoretical predictions for $\sigma$ are simple polynomial functions of $\pi$, which we compute as explained in the previous section. However, the predictions rely on the knowledge of the PDFs and of the FFs, which is imperfect. The rest of this section describes how the corresponding parametric uncertainties are included in the fit by introducing nuisance parameters that we collectively denote as `$\nu$'.

We employ Hessian representations of the PDFs and of the FFs, which consist of a set of functions---interpolated on a grid---that we generically denote as
\begin{equation}
    f_\alpha^{(k)}(\xi,Q^2)\,,
\end{equation}
where $k=1,\ldots,N$ labels the $N$ members of the set of functions, while `$\alpha$' runs over the different partons. One additional function, $f_\alpha^{(0)}(\xi,Q^2)$, is the `central value' PDF or FF. In the case of PDFs, the first argument $\xi$ is interpreted as the Bjorken $x$, while it corresponds to the $z$ variable in the case of the FFs. 

The Hessian set defines a basis for the PDFs or FFs, which reflects the current experimental knowledge of these functions. The central value is the best fit to current data and the members account for the uncertainties. In the particularly simple case of a linear Hessian set, the parametrization is 
\begin{equation}\label{eq:nupar}
    f_\alpha(\xi,Q^2;\nu)
    =f_\alpha^{(0)}(\xi,Q^2)+
    \sum\limits_{k=1}^{N}
    \nu_k\left[f_\alpha^{(k)}(\xi,Q^2)-f_\alpha^{(0)}(\xi,Q^2)\right].
\end{equation}
The nuisance parameters $\nu$ can be effectively interpreted as the parameters that are measured in the PDF fit. They are defined such that their best-fit value is $\nu=0$, they are uncorrelated and have unit variance.

The theoretical predictions for the binned cross sections $\sigma$ depend on the PDFs and on the FFs and so in turn they depend on $\nu$ through Eq.~(\ref{eq:nupar}) and not only on the parameters of interest $\pi$. Namely, $\sigma=\sigma(\pi,\nu)$. The $\chi^2$ (i.e., the log likelihood ratio) associated to the measurement of the observables ${\mathcal{O}}[\sigma]$ and to the prior determination of the PDFs, of the FFs and of the parameters of interest, thus reads
\begin{equation}\label{eq:chi2}
    \chi^2(\pi,\nu)=\left[{\mathcal{O}}[\sigma(\pi,\nu)]-\hat{\mathcal{O}}\right]^t\Sigma_{\rm{st}}^{-1}\left[{\mathcal{O}}[\sigma(\pi,\nu)]-\hat{\mathcal{O}}\right]+\sum\limits_k\nu_k^2+\pi^t\Sigma_{{\textsc{pdg}}}^{-1}\pi\,.
\end{equation}
In the equation, $\hat{\mathcal{O}}$ denotes the measured value of the observables, which we take to coincide with the theoretically predicted value when the parameters of interest are at the PDG best-fit (reported in Table~\ref{tab:CKMPDG}) and for central value nuisance parameters, namely
\begin{equation}
    \hat{\mathcal{O}}={\mathcal{O}}[\sigma(\pi_{\textsc{pdg}},0)]\,.
\end{equation}
The covariance matrix associated with the statistical error on the measurement of ${\mathcal{O}}$, $\Sigma_{\rm{st}}$, is obtained by error propagation as explained in Section~\ref{subsec:OBS}. The covariance matrix of the current determination of the parameters of interest, $\Sigma_{{\textsc{pdg}}}$, is obtained from the PDG errors in Table~\ref{tab:CKMPDG} ignoring correlations.

In general, the nuisance parameters have a relatively small effect on the PDFs and FFs, with the remarkable exception (see, e.g.~\cite{PDF4LHCWorkingGroup:2022cjn}) of the PDFs near $x=1$ where the PDF uncertainties exceed several tens or 100\,$\%$. The PDFs modelling offered by Eq.~(\ref{eq:nupar}) becomes unreliable in this region, which is thus excluded from the analysis as previously explained. Since the nuisance parameters effects are small in the kinematic region of interest, they can be treated at the linear order by expanding the cross section predictions as
\begin{equation}\label{eq:XSnu}
    \sigma(\pi,\nu)\simeq \sigma(\pi,0)+\sum_k\nu_k^{\textsc{p}}\Delta^{k}_{\textsc{p}}+\sum_k\nu_k^{\textsc{f}}\Delta^{k}_{\textsc{f}}\,,
\end{equation}
where the two sums run over the nuisance parameters $\nu_k^{\textsc{p}}$ associated with the PDFs and to the parameters $\nu_k^{\textsc{f}}$ associated with the FFs for bottom and charm tagging. 

The PDF variation terms $\Delta^{k}_{\textsc{p}}$ are determined by Monte Carlo integration in each bin similarly to what described in Section~\ref{subsec:OBS} for the binned cross sections calculation, but employing the PDF variation of the $k$-th member---namely, the difference $f^{(k)}_\alpha-f^{(0)}_\alpha$---in place of the PDFs inside the cross section formulas. This leads to smaller Monte Carlo errors than integrating the formulas with the $k$-th PDF member and subtracting the central value integral. The FFs are introduced (see again Section~\ref{subsec:OBS}) through $Q^2$-dependent efficiency/mistag factors. The FFs are set to the central value for the determination of these factors in the PDF variation terms $\Delta^{k}_{\textsc{p}}$. Conversely, the FFs are varied and the PDFs are kept to the central value for the calculation of the FF variation terms $\Delta^{k}_{\textsc{p}}$. Our calculation strategy automatically delivers cross sections as a function of the parameters of interest. However, it is an excellent approximation to set them to the PDG best-fit ($\pi=\pi_{\textsc{pdg}}$) in the nuisance variation terms $\Delta^{k}_{\textsc{p}}$ and $\Delta^{k}_{\textsc{f}}$.

Starting from Eq.~(\ref{eq:XSnu}), the linear expansion in $\nu$ of the observables ${\mathcal{O}}$ is obtained as a Taylor series and it is used in Eq.~(\ref{eq:chi2}) to obtain a $\chi^2$ that is quadratic in $\nu$, enabling the analytical treatment of nuisance profiling.\,\footnote{\label{fn:NL}The $D^*$-meson FFs Hessian set that we employ, from~\cite{Anderle:2017cgl}, is a quadratic and not a linear set and the $\pm1\,\sigma$ variations associated with the last nuisance parameter are not even approximately symmetric. The corresponding quadratic term is included in Eq.~(\ref{eq:XSnu}) and the exact dependence on the nuisance parameter is retained in the observables. Profiling over this single non-linear nuisance parameter is performed numerically.}\,Actually, the predictions can be linearised also with respect to 
the parameters of interest---by expanding around $\pi=\pi_{\textsc{pdg}}$---to an excellent approximation.
This makes the $\chi^2$ a quadratic form both in $\nu$ and in the relative variation from the PDG value of the parameters of interest, $\delta \pi$, which is defined as $\delta \pi\equiv(\pi-\pi_{\textsc{pdg}})/\pi_{\textsc{pdg}}$.

The Hessian PDFs and FFs sets that we employ in our analysis are accurately described in Appendix~\ref{app:pdf_ff}. The choice of the set, and in particular of the PDF set, plays a significant role the analysis because it sets the degree of flexibility of the PDFs parametrization~(\ref{eq:nupar}). It turns out (see Section~\ref{subsec:Fit-Anatomy} and Figure~\ref{fig:pdf_basis}) that the default 40 members Hessian set from PDF4LHC21~\cite{PDF4LHCWorkingGroup:2022cjn} Hessian set is not sufficiently expressive for our purposes and it is not used in the analysis. We employ instead a custom 300 members set that we obtain from the PDF4LHC21 constituents Monte Carlo set as explained in Appendix~\ref{app:pdf_ff}. 

\section{Results}
\label{sec:CKMd}

In this section we collect our main results. We present in Section~\ref{subsec:FitRes} the sensitivity projections on the CKM elements of our baseline fit configuration and we discuss the role played by the different measurements. The robustness of the results is assessed in Section~\ref{subsec:Fit-Anatomy} by studying variants of the baseline analysis. In Section~\ref{sec:PDFdet} we report our assessment of the perspectives for PDF measurements at the $\nuMuC$, which we obtain by studying the determination of the PDF nuisance parameters in our fit. Finally, in Section~\ref{sec:3TeVres} we quantify the CKM sensitivity achievable at a first stage MuC with a centre of mass energy of 3\,TeV.

\subsection{CKM matrix elements determination}
\label{subsec:FitRes}

\begin{table}[t]
    \centering
\begin{tabular}{|c|c|c|c|c|c|}
\hline
 Relative & \multirow{2}{*}{PDG} & \multirow{2}{*}{CC} & \multirow{2}{*}{(CC+NC)$_{\rm Stat}$} & ${\nuMuC}_{10}$
 & Gain \\
Precision $[\%]$ & \ & \ & \ & (CC+NC) &
$\textrm{PDG}/{\nuMuC}_{10}$
\\
\hline
 $ |V_{ud}|$ & 0.033 & 0.032  & 0.0019 & 0.0032  & 10  \\ \hdashline
 $ |V_{us}|$ & 0.38  & 0.19   & 0.017  & 0.067  & 5.6  \\ \hdashline
 $ |V_{cd}|$ & 1.8   & 1.6   & 0.0082 & 0.041    & 44 \\ \hdashline
 $ |V_{cs}|$ & 0.62  & 0.24   & 0.0033 & 0.016   &  39\\ \hdashline
 $ |V_{cb}|$ & 2.9   & 0.63   & 0.19   & 0.24    & 12 \\ \hdashline
 $ |V_{ub}|$ & 5.2   & 1.6   & 1.5    & 1.6    &  3.3\\
 \hline
 $ \sin \theta_{\textsc{w}}$      & 0.017 & --    & 0.0095  & 0.015  & 1.1 \\
\hline
\end{tabular}
    \caption{Relative global precision, in \%, on the absolute values of CKM elements and on $\sin \theta_{\textsc{w}}$ from the PDG~\cite{ParticleDataGroup:2024cfk} and after adding measurements at the 10\,TeV $\nuMuC$ in the baseline fit configuration and with high (1500\,ab$^{-1}$) integrated luminosity. Only charged-current observables are included for the CC column. Parametric uncertainties are ignored in the (CC+NC)$_{\rm Stat}$ column.\label{tab:global_sensitivity_fit}}
\vspace{-3mm}
\end{table}

Our baseline fit configuration exploits the binned cross section ratios listed in Table~\ref{tab:Oblist} and it assumes a tungsten nucleon target as in Table~\ref{eq:isoN}, with the high-luminosity setup in Eq.~\eqref{eq:HighL}. The parametric uncertainties from PDFs and FFs are included as explained in Section~\ref{subsec:Parametric_Uncertainties} and Appendix~\ref{app:pdf_ff} using a 300 members Hessian set derived from the PDF4LHC21~\cite{PDF4LHCWorkingGroup:2022cjn} constituents set. The global $68\%$ relative precision on each parameter of interest---obtained by profiling the $\chi^2$~(\ref{eq:chi2}) both over the nuisance parameters and over the other parameters of interest---is reported on the fifth column of Table~\ref{tab:global_sensitivity_fit}. The last column of the table shows the precision gain, which is defined as the ratio between the present-day PDG uncertainty (second column) and the result of our baseline fit. The third column of the table displays the result of a fit that only exploits the CC cross section measurements, i.e.\,the top set of observables in Table~\ref{tab:Oblist}. Finally, the fourth column of the table reports the purely statistical sensitivity to the parameters of interest. They are obtained by setting to zero the nuisance parameters in the $\chi^2$, Eq.~\eqref{eq:chi2}, hence effectively ignoring the parametric uncertainties.

The baseline fit significantly improves upon the PDG uncertainties for all CKM elements. The precision on $|V_{ud}|$ improves by a factor of 10, reaching $3.2 \times 10^{-5}$, not far from the statistical reach. The precision on $|V_{us}|$, $|V_{cd}|$, and $|V_{cs}|$ reaches the $10^{-4}$ level, improving current uncertainties by a factor of 5.6, 44, and 39, respectively. Nevertheless, the precision reach for these elements is still driven by parametric uncertainties and quite far from the purely statistical one. Per-mille level precision is obtained for $|V_{cb}|$, an improvement of one order of magnitude with respect to today, while the uncertainty on $|V_{ub}|$ is at the one percent level, improving the current one by a factor of 3. The global precision on $|V_{cb}|$ and $|V_{ub}|$ is close to the purely statistical one, being limited by the relatively small number of $\nu_\mu \bar{c} \to \mu^- \bar{b}$, $\bar{\nu}_e c \to e^+ b$, and $\bar{\nu}_e u \to e^+ b$ partonic events.

By comparing the sensitivity reach in the baseline fit with the one obtained by including only CC processes (third column of Table~\ref{tab:global_sensitivity_fit}), we see a significant degradation for all CKM matrix elements (except $V_{ub}$). This gives a measure of the importance of the NC observables. While insensitive to the CKM, the NC measurements play a vital indirect role in their determination because they help constraining the nuisance parameters associated with the PDF and FF uncertainties.

\begin{figure}[t]
\centering
    \includegraphics[width=0.5\textwidth]{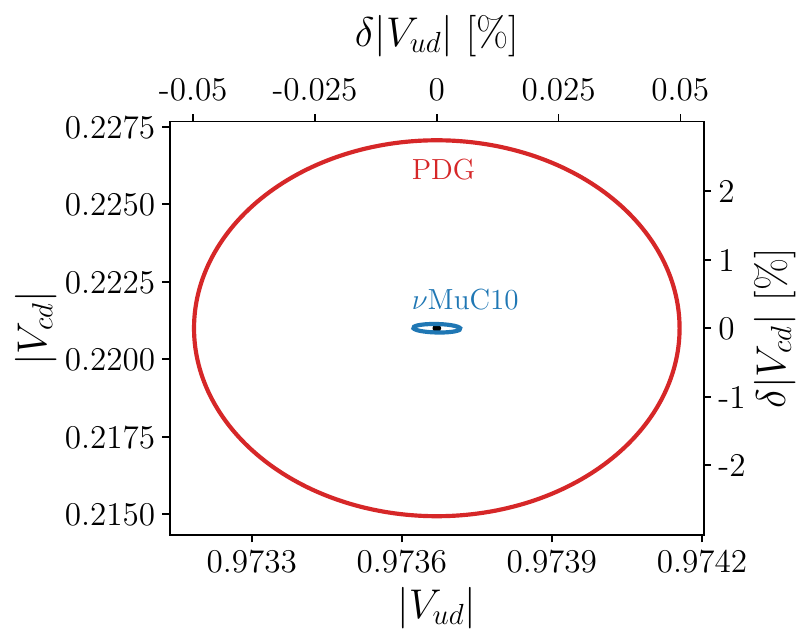}%
    \hfill
    \includegraphics[width=0.5\textwidth]{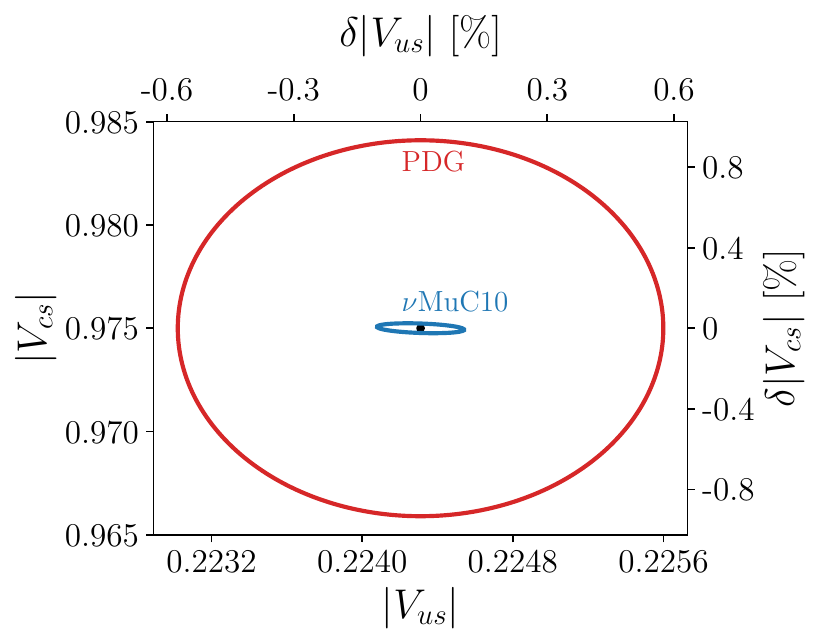} \\

    \includegraphics[trim={0 -15pt 0 0},clip,width=0.5\textwidth]{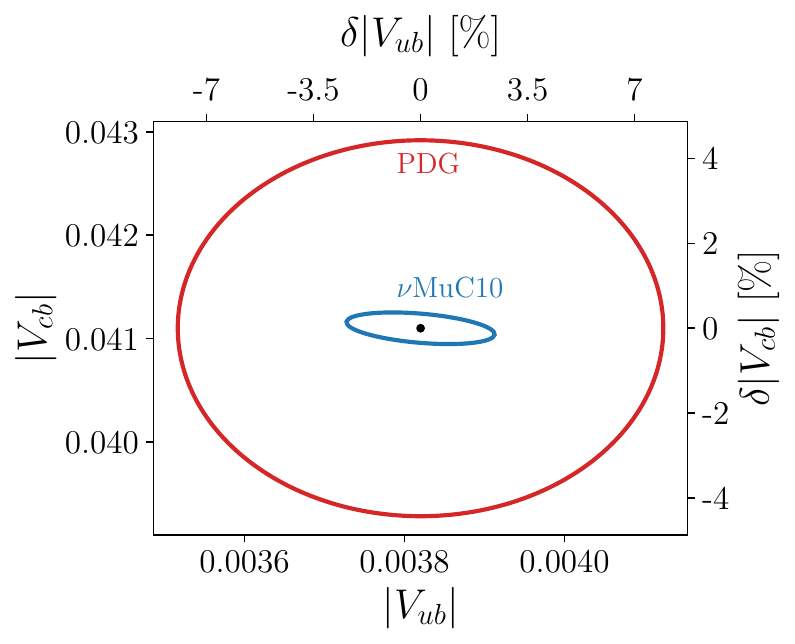}
    \hfill    \includegraphics[trim={0 -20pt 0 0},clip,width=0.47\linewidth]{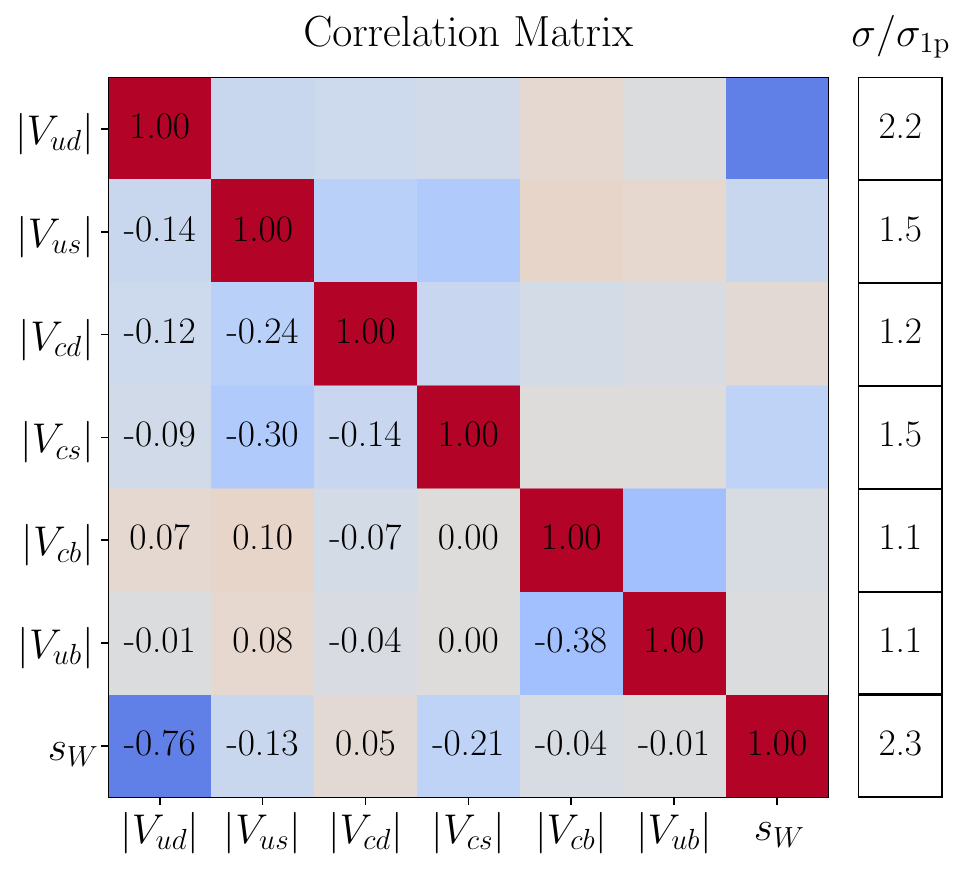} \\

    \caption{68\%\,CL ellipses for CKM pairs, showing the PDG prior (red) and the baseline fit (blue). The bottom-right panel shows the correlation matrix of the complete fit and the ratio $\sigma/\sigma_{\rm{1p}}$ between the 1-parameter and the global precision.
    }
    \label{fig:ckm_fit_ellipses}
\end{figure}

The first three panels of Figure~\ref{fig:ckm_fit_ellipses} offer a two-dimensional visualisation of the improvement in the CKM matrix elements determination. The plots are 68\%\,CL ellipses for selected pairs of CKM matrix elements, obtained by profiling over the other parameters of interest and over the nuisance parameters.
The starting point is the PDG, shown in red, while the ${\nuMuC}_{10}$ baseline fit results are shown in blue. 

The bottom-right panel of Figure~\ref{fig:ckm_fit_ellipses} shows the correlation matrix among the 7 parameters of interest obtained in the baseline fit. It also reports another indicator of correlation---labelled as the ratio $\sigma/\sigma_{\rm{1p}}$---for each parameter. Here $\sigma$ denotes the global 68\%\,CL precision while $\sigma_{\rm{1p}}$ is the precision of a single-parameter fit performed under the assumption that all the other parameters of interest are equal to the PDG central value. It is obtained by marginalizing the $\chi^2$ over the nuisance parameters, but setting the other parameters of interest to their PDG central value rather than profiling over them as is done instead for the determination of the global precision. The correlations are generically small, except for a $-76\,\%$ correlation between $V_{ud}$ and $s_W$ which also reflects in a ratio $\sigma/\sigma_{\rm{1p}}$ above 2 for these two parameters. Should an independent determination of the Weinberg angle with less than $10^{-4}$ precision become available---as expected for instance from FCCee~\cite{FCC:2025lpp}---one could fix this parameter in our fit. In this case, the global precision on $V_{ud}$ would improve to $2.1 \times 10^{-5}$, while the sensitivity on the other CKM elements remains unchanged.

\begin{figure}
    \centering    \includegraphics[width=0.65\linewidth]{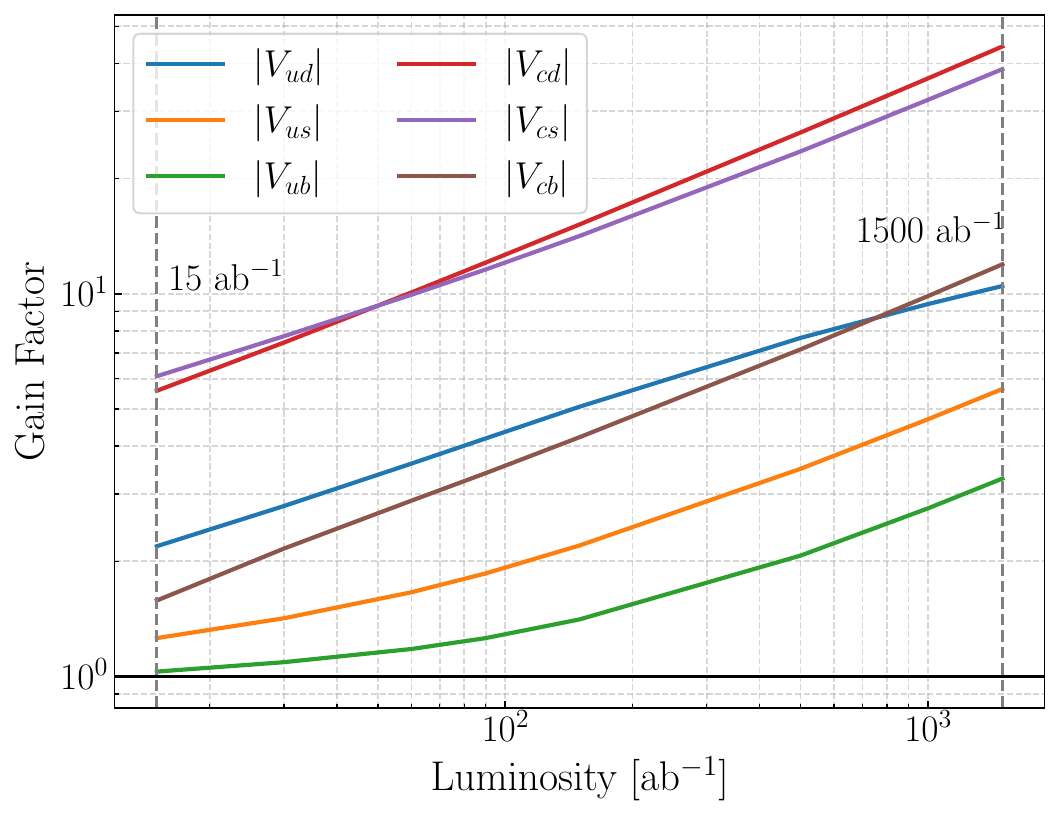}
    \caption{The gain---relative to current knowledge---in the CKM parameters precision as a function of the integrated luminosity at the 10\,TeV $\nuMuC$ experiment.
    }
    \label{fig:luminosity-gainy}
\end{figure}

Such advance in the knowledge of the CKM matrix is a strong physics case for the $\nuMuC$ experiment, which however relies on a placement of the experiment and on a design of the target---see Section~\ref{ssec:ExS}---that enables to collect the `high' integrated luminosity of 1500\,ab$^{-1}$ as in Eq.~(\ref{eq:HighL}). Figure~\ref{fig:luminosity-gainy} shows the effect on the gain factor---defined as before as the ratio between the PDG error and the global precision on each CKM element---of lowering the integrated luminosity down to the `low' luminosity setup with 15\,ab$^{-1}$, Eq.~\eqref{eq:LowL}. For a luminosity above around 100\,ab$^{-1}$ the precision scales with the square root of the luminosity\,\footnote{This behaviour sets in at much higher luminosity for $V_{ud}$, due to the correlation of this parameter with $\sin \theta_{\textsc{w}}$, whose determination in the fit is still driven by the prior.}\,as expected for a statistics- (and not systematics-) dominated fit, in spite of the dominant role played by parametric uncertainties for most parameters. This is because ours is the result of a combined determination of the parameters of interest and of the nuisance parameters that control the parametric uncertainties and the precision on the nuisance parameters determination also scales with $\sqrt{\mathcal{L}}$. For lower luminosity, the precision on some CKM elements saturates to the PDG prior and the gain approaches one. In the low luminosity scenario the gain is limited (below~2) for almost all the parameters, except for $V_{cd}$ and $V_{cs}$, for which it is about 4 and 6, respectively. The perspective of a radical improvement in the CKM parameters determination is thus a strong physics case for the design of a high-luminosity $\nuMuC$ experiment.

\subsubsection*{Impact of the SIDIS processes}

\begin{table}[t]
    \centering
\begin{tabular}{|c|c|c|c|c|c|c|}
\hline
 Relative & \multirow{2}{*}{PDG}& DIS & DIS  & DIS & DIS & DIS \\
Precision $[\%]$ &\ & only & + $b$-tags \ & + $b,c$-tags & + $b,s$-tags  &  + $b,c,s$-tags \\
\hline
 $ |V_{ud}|$ & 0.033 & 0.028 & 0.027   & 0.026 & 0.017 & 0.0032   \\ \hdashline
 $ |V_{us}|$ & 0.38 & 0.37  & 0.37   & 0.36  & 0.11 & 0.067    \\ \hdashline
 $ |V_{cd}|$ & 1.8 & 0.90   & 0.88   & 0.84 & 0.57 & 0.041     \\ \hdashline
 $ |V_{cs}|$ & 0.62 & 0.037  & 0.030   & 0.024 & 0.021 & 0.016  \\ \hdashline
 $ |V_{cb}|$ & 2.9 & 2.9  & 0.81    & 0.25  & 0.27 & 0.24   \\ \hdashline
 $ |V_{ub}|$ & 5.2 & 5.2   & 2.1    &  1.6  & 1.6  & 1.6   \\
 \hline
\end{tabular}
    \caption{Relative global precision on the CKM elements at the 10\,TeV $\nuMuC$ with high (1500\,ab$^{-1}$) integrated luminosity. Starting from the current PDG uncertainties, the columns from left to right correspond to: only inclusive DIS; DIS plus bottom-tagged SIDIS; DIS plus bottom- and charm-tagged SIDIS; DIS plus bottom- and strange-tagged SIDIS. The last columm is the baseline fit configuration with all measurements included.
    }

    \label{tab:fit_different_tags}
\end{table}

\begin{figure}[t]
\centering
    \includegraphics[width=0.48\textwidth]{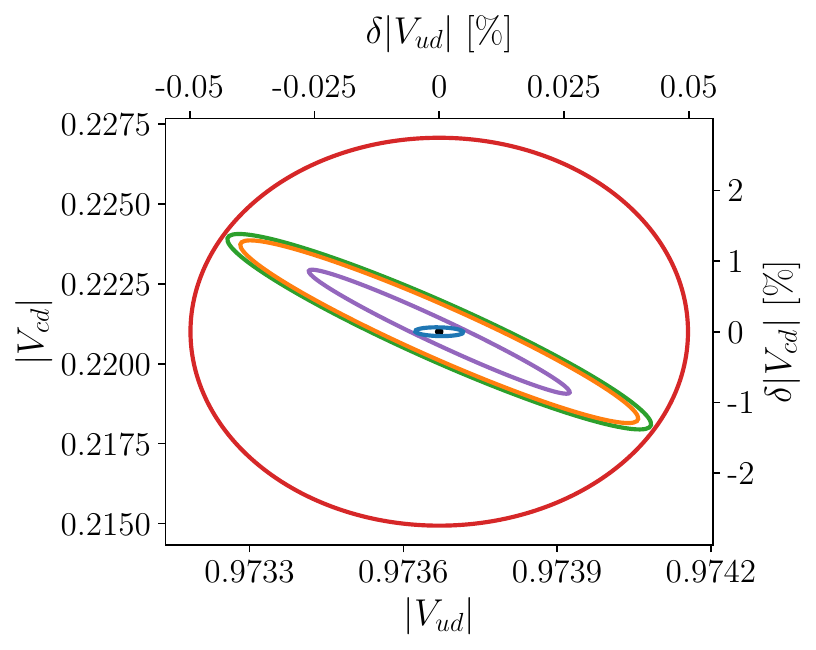}%
    \hfill
    \includegraphics[width=0.51\textwidth]{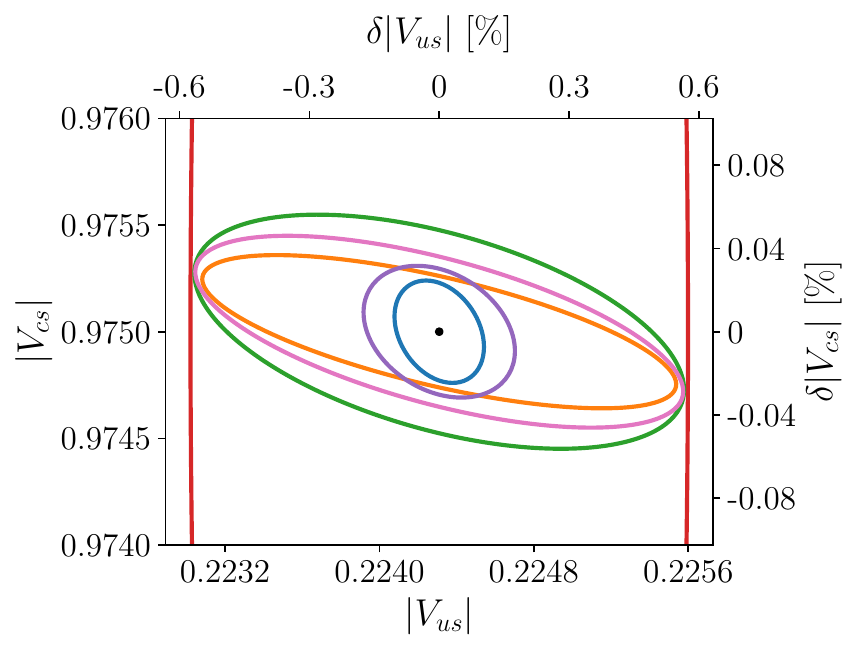} \\
    \includegraphics[width=0.48\textwidth]{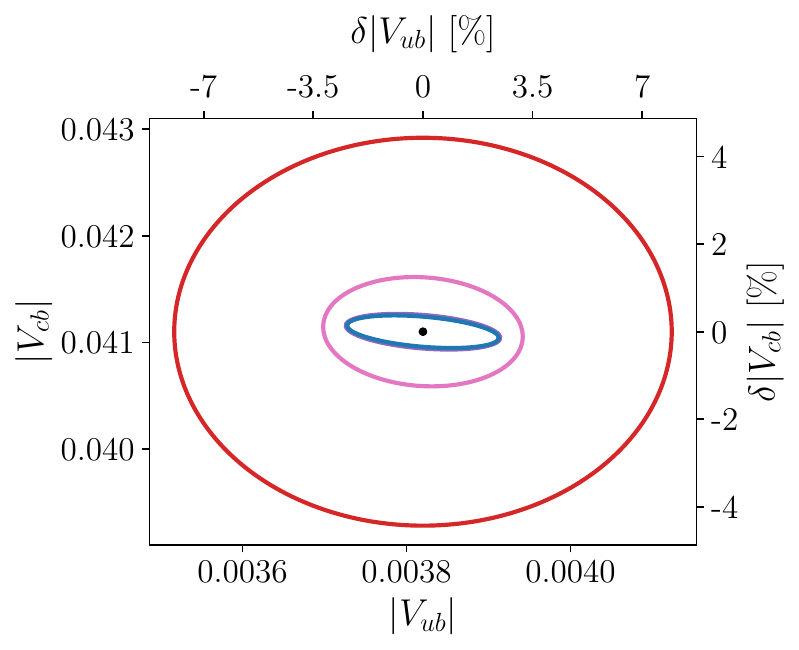}
    \includegraphics[width=0.47\textwidth]{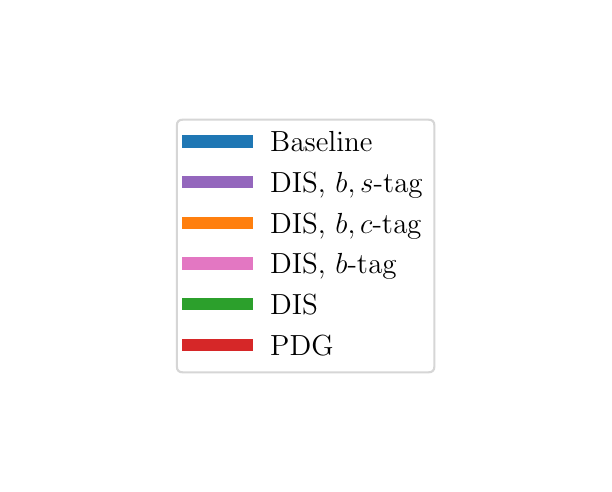}\\
    
    \caption{68\%\,CL ellipses for CKM pairs when only a partial set of observables is included as detailed in the legend.The PDG prior is shown in red and the baseline fit in blue like in Figure~\ref{fig:ckm_fit_ellipses}.}
    \label{fig:ckm_fit_ellipses_observables}
\end{figure}

Table~\ref{tab:fit_different_tags} and Figure~\ref{fig:ckm_fit_ellipses_observables} show the impact of the different SIDIS measurements on the sensitivity to each CKM matrix element. The third column reports the results of a fit which includes only DIS cross sections, specifically the ratios $R^{\mu / e}$ and $R^{\textsc{CC}/\textsc{NC}}$ (see definitions in Table~\ref{tab:Oblist}). These observables, alone, are sufficient to reach a precision on $V_{cs}$ at the $10^{-4}$ level, already close to the one of the baseline fit (reported again for convenience in the last column). This can also be seen from the green 68\% CL ellipse in the top-right panel of Figure~\ref{fig:ckm_fit_ellipses_observables}. While the global precision on the other CKM matrix elements remains at the level of the PDG prior, the DIS ratios strongly constrain one direction in the $V_{ud}$-$V_{cd}$ plane, as can be seen from the green ellipse in the top-left panel of the same figure.

The results shown in the fourth column are obtained by including also the bottom-tagging ratios $R^{ib}$ (where $i=e,\mu,\nu$) in the fit. As expected, this mainly impacts $V_{cb}$ and $V_{ub}$, whose precision improves by factors of about 3.5 and 2.5, respectively. The precision on $V_{cs}$ also improves, although only by approximately 20\%.  In Figure~\ref{fig:ckm_fit_ellipses_observables}, this setup with only DIS and bottom-tagged SIDIS is displayed as a pink ellipse. It is not shown in the top-left panel because the corresponding ellipse overlaps completely with the previous DIS-only result (green curve).

Adding also charm-tagged observables to the fit improves (see the fifth column of the table) the precision of $V_{cb}$ by another factor of 3 and the ones on $V_{ub}$ and $V_{cs}$ by about 20\%. The effect on $V_{cd}$ is instead minor. The lines in Figure~\ref{fig:ckm_fit_ellipses_observables} corresponding to this setup are displayed in orange. In the bottom panel, this fit configuration is not shown because the corresponding ellipse overlaps with the blue one of the baseline fit.

The sixth column of Table~\ref{tab:fit_different_tags} corresponds to a fit in which we exclude the charm-tagged observables but include both bottom and strange tagging. The corresponding 68\% CL ellipses are displayed in purple in Figure~\ref{fig:ckm_fit_ellipses_observables}. In comparison with charm tagging, strange tagging produces stronger constraints on $V_{ud}$, $V_{us}$, $V_{cd}$, and $V_{cs}$. The impact on $V_{cb}$ and $V_{ub}$ is instead similar to the one of the charm tagging and the purple ellipse almost overlaps with the blue one of the baseline fit.

Finally, the last column and the blue ellipses reports the results of the baseline fit, which include all the observables listed in Table~\ref{tab:Oblist}. The combination of charm and strange tagging allows to lift the flat directions and strongly improve the precision on $V_{ud}$ and $V_{cd}$.

It is worth mentioning that in order to obtain these results it necessary to include the $s$-tagged observable $R^{e s}$ and the $\bar{s}$-tagged observable $R^{\mu \bar{s}}$, in spite of the fact that the corresponding cross sections do not receive contribution from the production of the quark (the $s$ and the $\bar{s}$, respectively) that the corresponding taggers are supposed to select. These cross sections only receive contribution from mistagging, but nevertheless are large (see Table~\ref{tab:XSlist}) because the mistag rates are large. The observable ratios $R^{e s}$ and $R^{\mu \bar{s}}$ are thus measured with good precision and they play an essential role in the global sensitivity by lifting otherwise poorly constrained directions in the parameter space.

\subsubsection*{SM mixing parameters determination}

\begin{table}[t]
    \centering
    \begin{tabular}{|c|c|c|c|c|c|c|}
    \hline
     \multirow{2}{*}{PoI} & \multirow{2}{*}{PDG} & \multicolumn{2}{c|}{Relative Precision [\%]} & Gain\\
     \cline{3-4}
&   & PDG & ${\nuMuC}_{10}$ & PDG$/{\nuMuC}_{10}$ \\
     \hline
     $\sin \theta_{12}$ & $0.22500 \pm 0.00067$ & 0.30 & 0.028 & 11 \\
     $\sin \theta_{23}$ & $0.04185 \pm 0.00080$ & 1.9 & 0.23 & 8 \\
     $\sin \theta_{13}$ & $(3.69 \pm 0.11)\times 10^{-3}$ & 3.0 & 1.5 & 2 \\
     $\delta_{\textsc{cp}}$ & $1.144 \pm 0.027$ & 2.4 & 2.4 & 1 \\
     \hline
     $s_W^2$ & $0.23129 \pm 0.00004$ & 0.017 & 0.0068 & 2.5 \\ 
     \hline
    \end{tabular}
    \caption{PDG averages \cite{ParticleDataGroup:2024cfk} for the Parameters of Interest (PoI) of our baseline fit to the SM mixing angles and phase, the corresponding relative precisions and the global precision of the $\nuMuC$ experiment with 1500~ab$^{-1}$ at the 10\,TeV MuC. The last column reports the gain in precision.}
    \label{tab:global_sensitivity_fit_unitaryCKM}
\end{table}

In the SM, the CKM matrix is unitary and its elements can be expressed in terms of 3 angles ($\theta_{12}$, $\theta_{23}$ and $\theta_{13}$) and one phase ($\delta_{\textsc{cp}}$) using a standard parametrization (see e.g.\,Ref.~\cite{ParticleDataGroup:2024cfk}). The 2024 PDG average for these parameters is reported in Table~\ref{tab:global_sensitivity_fit_unitaryCKM}. Our measurements are sensitive to the absolute value of 6 CKM elements that, to an excellent approximation, read
\begin{equation}
\begin{split}
    |V_{ud}| &= c_{12} c_{13}\,,\quad
    |V_{us}| = s_{12} c_{13}\,,\quad
    |V_{ub}| = s_{13}\,,\quad
    |V_{cb}| = s_{23} c_{13}\,,\\
    |V_{cd}| &\approx s_{12} c_{23} + c_{12} s_{13} s_{23} \cos \delta_{\textsc{cp}} \,,\quad
    |V_{cs}| \approx c_{12} c_{23} - s_{12} s_{13} s_{23} \cos \delta_{\textsc{cp}} \,,
\end{split}
\label{eq:CKM_unitarity_rel}
\end{equation}
where $s_{ij} = \sin \theta_{ij}$ and $c_{ij} = \cos \theta_{ij}$.

In a unitary parametrization such as the one in Eq.~(\ref{eq:CKM_unitarity_rel}), the trigonometric relations among the elements define a sub-space of the full $|V_{ij}|$ parameter space that is instead considered in our baseline fit that does not enforce CKM unitarity. Therefore the accuracy in the SM mixing parameters cannot be rigorously extracted from the global fit precision because this sub-space might not be aligned with the correlations in the $|V_{ij}|$ space. A dedicated fit is needed, using Eq.~(\ref{eq:CKM_unitarity_rel}) in the $\chi^2$~(\ref{eq:chi2}) and treating the SM mixing angles $\theta_{ij}$ and $\delta_{\textsc{cp}}$ as parameters of interest. The results are reported in Table~\ref{tab:global_sensitivity_fit_unitaryCKM}. 

The measurement of the Cabibbo angle, $\theta_{12}$, emerges from the combination of the measurements of $|V_{us}|$, $|V_{ud}|$, $|V_{cd}|$, and $|V_{cs}|$. The improvement from the PDG is about a factor of 11. Given the smallness of $\theta_{13}$, the parameters $s_{13}$ and $s_{23}$ are---see Eq.~(\ref{eq:CKM_unitarity_rel})---essentially in one-to-one correspondence with $|V_{ub}|$ and $|V_{cb}|$ and these two parameters are correlated only among themselves (with about $-40\%$ correlation) in the baseline fit. Therefore the precision on $s_{13}$ and $s_{23}$ (and the correlation) of the SM fit in Table~\ref{eq:CKM_unitarity_rel} is very similar to the non-unitary fit results. The gain factors from the PDG are about 8 and 2, respectively, smaller than the gains in the baseline fit shown in Table~\ref{tab:global_sensitivity_fit}. This is due to a better precision on the CKM angles in the PDG combination once unitarity relations are imposed, as can be seen by comparing the PDG relative precision in the two tables.

Finally, the sensitivity on the Weinberg angle improves by a factor of about 2.5 from the current precision. The improvement relative to the non-unitary fit is mainly due to the correlation between the Weinberg angle and $|V_{ud}|$, about $-76\%$. In the SM fit, the correlation between $s_{12}$ and $s_W$ reduces to $+24\%$.

In line of principle, our determination of the absolute value of $6$ CKM elements is sufficient for the measurement of all the 4 SM mixing parameters, including $\delta_{\textsc{cp}}$. This occurs as a consequence of unitarity in spite of the fact that our observables are even under the CP symmetry. However, the dependence on $\delta_{\textsc{cp}}$ is---see Eq.~(\ref{eq:CKM_unitarity_rel})---strongly suppressed by $s_{13}\cdot s_{23}\sim10^{-4}$ and the $\nuMuC$ sensitivity to $\delta_{\textsc{cp}}$ is not competitive with current precision.

\subsection{Analysis variants}
\label{subsec:Fit-Anatomy}

In this section we check the robustness of the results under variations of the baseline analysis assumptions. Specifically, we study the dependence on the PDF Hessian set employed in the fit and on the composition of the target. 

\subsubsection*{Dependence on the PDF set}

The Hessian PDF set employed in the fit can have a considerable impact on the results, because of the following. The $\nuMuC$ DIS and SIDIS measurements are very accurate and they enable (see Section~\ref{sec:PDFdet}) a much better determination of the PDFs than the current data. This is not the case for the FFs, definitely because we did not exploit the $z$ variable distribution and we used instead a single bin in $z$. Standard PDF Hessian sets such as for instance PDF4LHC21\_40~\cite{PDF4LHCWorkingGroup:2022cjn} are designed to offer a compressed representation of the PDFs and their uncertainties that is faithful with present-day data, but it does not necessarily offer a parametrization that is adequate to fit more precise future data. In particular, PDF4LHC21\_40 is obtained by compressing the full PDF4LHC21 set, which consists of 900 Monte Carlo PDF replicas, in a Hessian set with only 40 members. The compression is based---see Appendix~\ref{app:pdf_ff}---on the truncation of the PDF covariance matrix to the subspace of its $N=40$ Eigenvectors with largest Eigenvalues. The choice $N=40$ defines the number of PDF members that in turn determines the degree of expressivity of the PDF parametrization in Eq.~(\ref{eq:nupar}). It is found to be an adequate choice for current data because the neglected Eigenvectors, with smaller Eigenvalues than the first $40$ ones, have completely negligible effect on any prediction in comparison with the present-day errors. On the contrary, future and more precise data might be sensitive to the neglected Eigenvectors. We should thus study the dependence of our results on the PDF set, and in particular on the number of Eigenvectors employed in the truncation.

\begin{figure}[t]
    \centering   
    \includegraphics[width=0.65\linewidth]{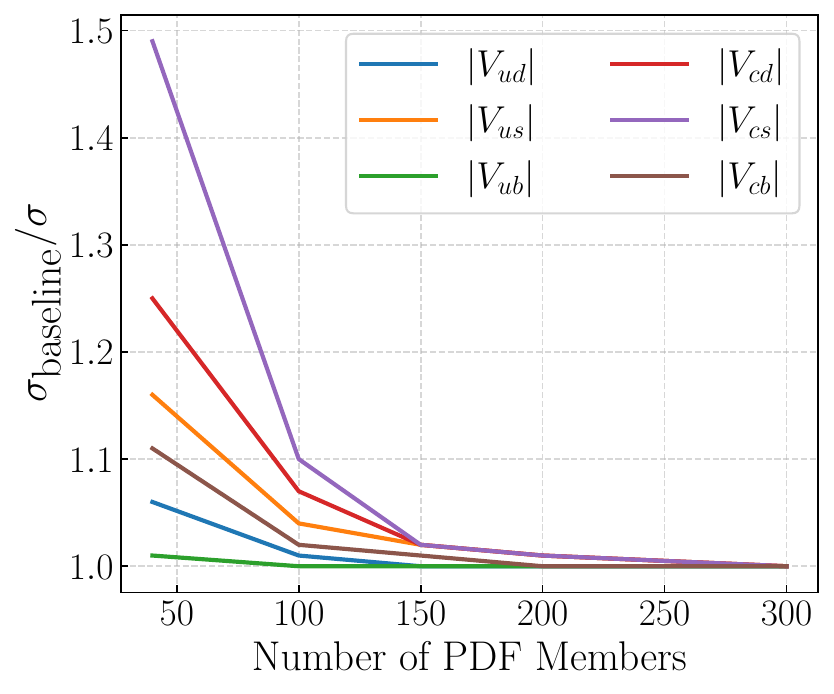}
    \caption{The CKM parameters global precision in the baseline analysis, with a 300 members PDF Hessian set, divided by the precision obtained by varying the number of members in the set. The results are for the $\nuMuC$ experiment at the 10\,TeV collider and with 1500\,{\textrm{ab}}$^{-1}$ integrated luminosity.}
    \label{fig:pdf_basis}
\end{figure}

Starting from the PDF4LHC21~\cite{PDF4LHCWorkingGroup:2022cjn} set of 900 Monte Carlo replicas, we construct new Hessian sets with a variable number of members as explained in Appendix~\ref{app:pdf_ff}. By repeating the fit with these new sets, we study the dependence of the global sensitivity to the CKM parameter of interests. The sensitivity gain relative to the baseline result, obtained with~300 replicas, is shown in Figure~\ref{fig:pdf_basis}. Starting from 40, which is the number of members chosen in the PDF4LHC21\_40 set, the sensitivity to $V_{ud}$ and to $V_{cd}$ (and to the other parameters, to a lesser extent) deteriorates appreciably when the number of members is increased above 40.  This is due to the enhanced expressivity of the PDF parametrization in the larger Hessian sets, which enables to reach new configurations for the PDFs that compensate more accurately the effects on the observables due to the variation from the central value of the CKM elements, hence reducing the global sensitivity. However, the sensitivity quickly saturates reaching a plateau above around 100 members and it remains very stable when the number of members is further increased. This suggests that our baseline results with 300 members do not suffer from unphysical correlations among the PDFs induced by a not sufficiently expressive PDF modelling.

\subsubsection*{Target nucleon dependence}

\begin{table}[t]
    \centering
\begin{tabular}{|c|c|c|c|c|c|}
\hline
 Relative Precision $[\%]$&  Proton & Neutron & Isoscalar & Silicon & Tungsten \\
\hline
 $ |V_{ud}|$  & 0.0034 & 0.0032 & 0.0030 & 0.0030 & 0.0032 \\ \hdashline
 $ |V_{us}|$   &0.049 &0.10 & 0.043 & 0.051  & 0.067 \\ \hdashline
 $ |V_{cd}|$ &0.059 &0.039 & 0.035 & 0.036  & 0.041  \\ \hdashline
 $ |V_{cs}|$  &0.016 &0.018 & 0.010 & 0.011  & 0.016 \\ \hdashline
 $ |V_{cb}|$    &0.24 &0.25 & 0.24 & 0.24   & 0.24 \\ \hdashline
 $ |V_{ub}|$    &1.3 &2.1 & 1.5 & 1.5   & 1.6 \\
\hline
\end{tabular}
\caption{Relative global precision on the absolute values of CKM elements at the 10\,TeV $\nuMuC$ and 1500\,ab$^{-1}$ of integrated luminosity for different target nucleons. 
\label{tab:diff-target-precision}}
\end{table}

There is at present no design for the $\nuMuC$ experiment and in particular not much is known on the possible composition of the target. The design of Ref.~\cite{King:1997dx}, performed long ago for a neutrino detector parasitic to a 250\,GeV muon ring, envisaged a silicon-based target. The detector material for similar experiment such as FASER$\nu$, FASER$\nu$2 and SND@LHC is instead mostly tungsten. In the absence of dedicated experimental studies, we based our baseline sensitivity projections on tungsten, whose effective PDFs are estimated based on isospin symmetry as in Table~\ref{eq:isoN}. The CKM parameters precision obtained with the other nucleons listed in the same table are reported in Table~\ref{tab:diff-target-precision}. The results do depend on the nucleon, as expected, but the performance gap across different nucleons is moderate. 

It is interesting to compare the results for silicon with the one for an isoscalar nucleon, taking into account that silicon is almost exactly isoscalar, with $A/Z=2.01$. Nevertheless the results are not identical in the two cases and the difference is significant for the $V_{us}$ parameter sensitivity. This is due to the fact that when $A/Z$ is exactly equal to 2 the PDF for the $u$ and $d$ and for the $\bar{u}$ and $\bar{d}$ quarks are exactly equal. This induces exact relations among observables that are not affected by PDF uncertainties and can produce better CKM sensitivity. Such relations are broken for the silicon nucleon, resulting in a slightly inferior sensitivity, even if $A/Z$ is so close to 2. This also suggests that our results for silicon should be taken with a grain of salt: the difference between the up and the down PDFs is suppressed by a factor of $1-2Z/A\simeq 5$\,\textperthousand~relative to the sum and the same suppression applies to the uncertainties on the PDFs. Isospin breaking effects in the physical silicon nucleus can be larger than that, invalidating our results. On the contrary, the baseline tungsten nucleon is very far from the isoscalar limit and this concern does not apply.

Once the target material will be determined in the experiment design, more accurate sensitivity projections will be performed considering the scattering on the entire nucleus and the corresponding nuclear PDFs. With the high precision of the $\nuMuC$ measurements, isospin breaking effects will be relevant and possibly the uncertainties on the nuclear PDFs associated with nuclear effects will emerge as a limiting factor for the determination of the proton and neutron PDFs. Notice however that these uncertainties do not affect the extraction of the CKM parameters because the predictions depend on the physical nucleon PDFs and for the purpose of measuring the CKM there is no need to relate them to the PDFs for protons and neutrons. The physical nuclear PDFs are what enters in the fit and they will be determined together with the CKM elements.

\subsection{PDF determination}\label{sec:PDFdet}

As a by-product of the analysis, we can estimate the $\nuMuC$ experiment potential for the determination of the PDFs, by proceeding as follows. After profiling the full $\chi^2$ in Eq.~(\ref{eq:chi2}) over the nuisance parameters $\nu_k^{\textsc{f}}$ that describe the uncertainties on the FFs and over the parameters of interest $\pi$, we get the $\chi^2$ as a function of the PDF nuisance parameters $\nu_k^{\textsc{p}}$, with $k$ running over the 300 members of our baseline (see Appendix~\ref{app:pdf_ff}) PDF set. The profiled $\chi^2$ is a quadratic form $(\nu_k^{\textsc{p}})^t H \nu_k^{\textsc{p}}$ with Hessian matrix $H$. We diagonalize the Hessian matrix as
\begin{equation}
    H_{kk^\prime}=\sum\limits_\lambda O_{k\lambda}\frac1{\sigma_\lambda^2} O_{k^\prime\lambda}\,,
\end{equation}
where $\sigma_\lambda^2$ are the 300 Eigenvalues of the inverse Hessian---i.e.\,the square of the global precision of the $\nu^{\textsc{p}}$ parameters measurement---and $O$ is the orthogonal matrix whose columns are the Eigenvectors of the Hessian. The set of 300 functions
\begin{equation}
    \varphi_\alpha^{(\lambda)}(x,Q^2)
    =f_\alpha^{(0)}(x,Q^2)+\sigma_\lambda\sum\limits_k
     O_{k\lambda}
    \left[f_\alpha^{(k)}(x,Q^2)-f_\alpha^{(0)}(x,Q^2)\right]
    \,,
\end{equation}
are the variation at $1\sigma$ in the PDF space---see Eq.~(\ref{eq:nupar})---along the direction of the Eigenvectors. The functions $\varphi$ define the post-fit Hessian set that describes the uncertainties on the PDFs after including the $\nuMuC$ projected data.

\begin{figure}[p]
\centering
    \includegraphics[width=0.47\textwidth]{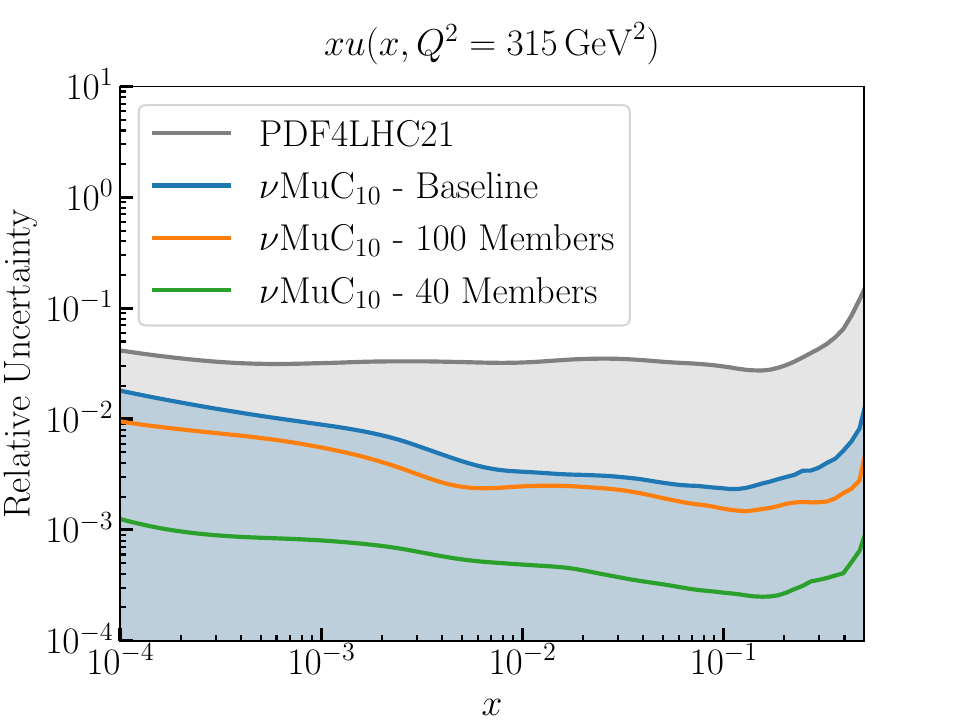} \hspace{-0.5cm}
    \includegraphics[width=0.47\textwidth]{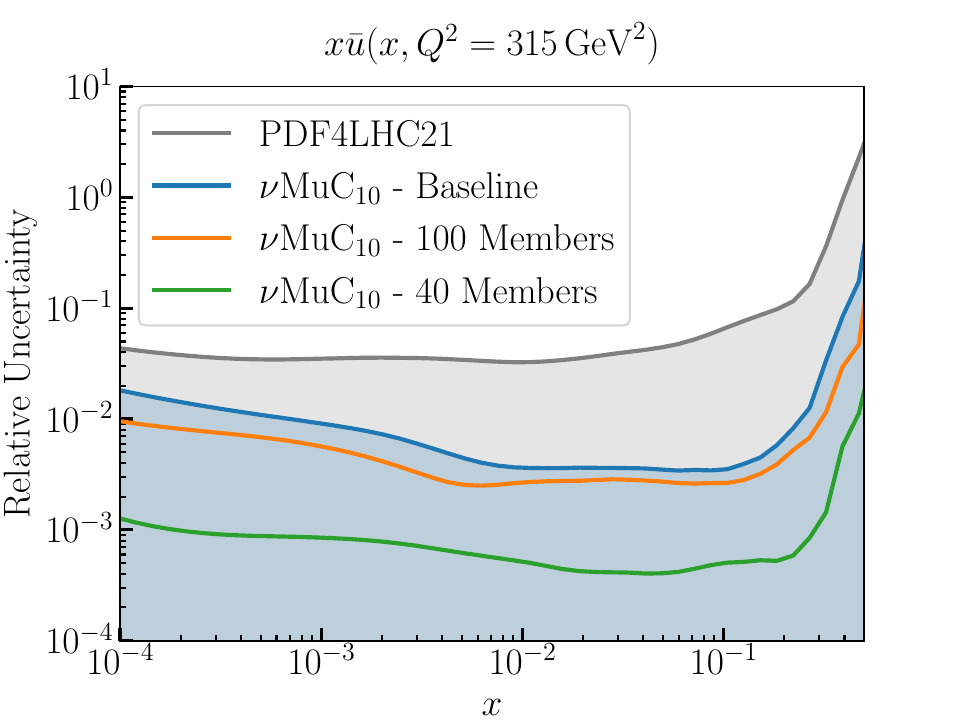}\\
    
    \includegraphics[width=0.47\textwidth]{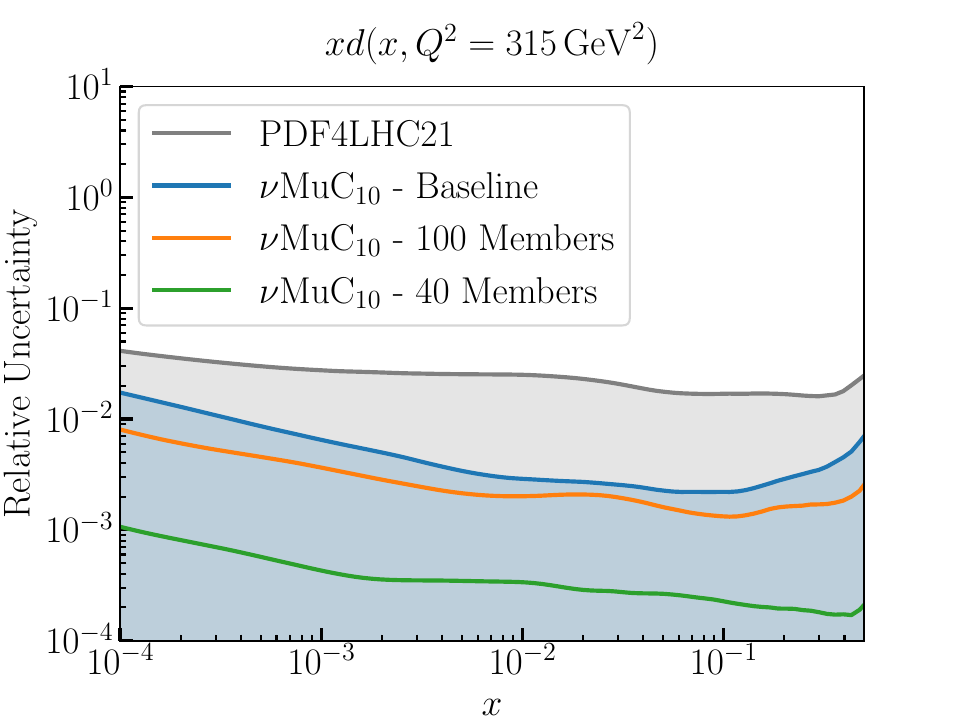} \hspace{-0.5cm}
    \includegraphics[width=0.47\textwidth]{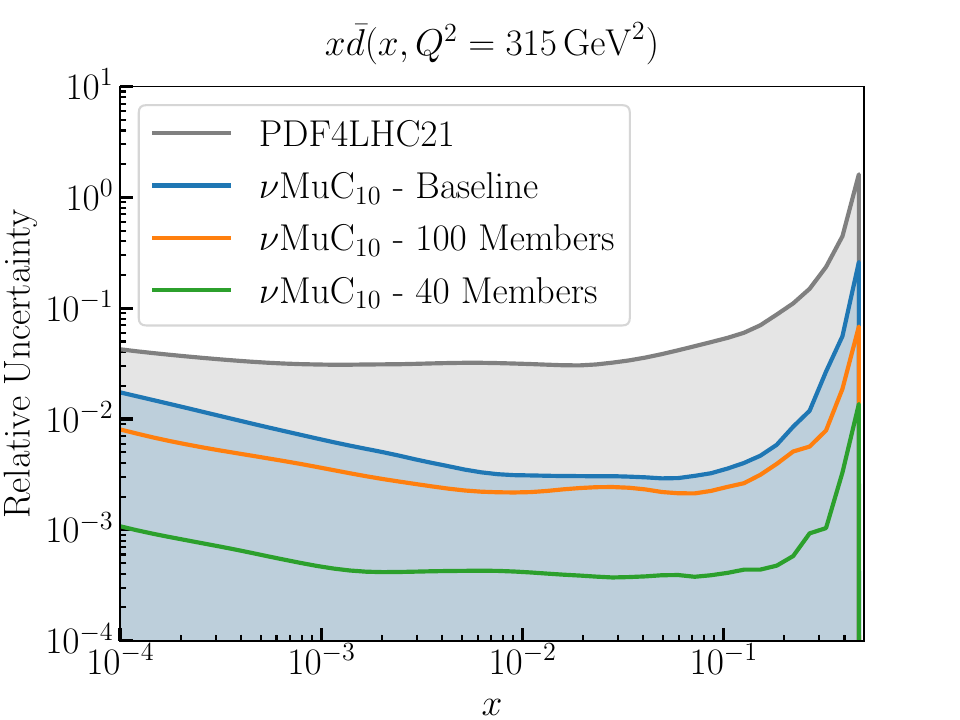}\\
    
    \includegraphics[width=0.47\textwidth]{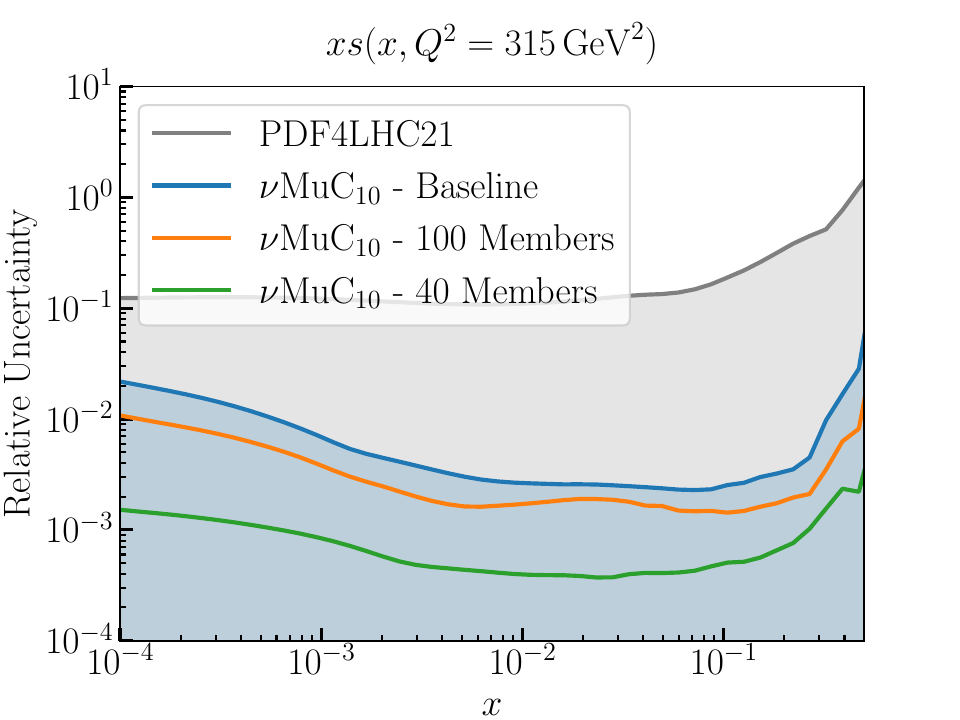} \hspace{-0.5cm}
    \includegraphics[width=0.47\textwidth]{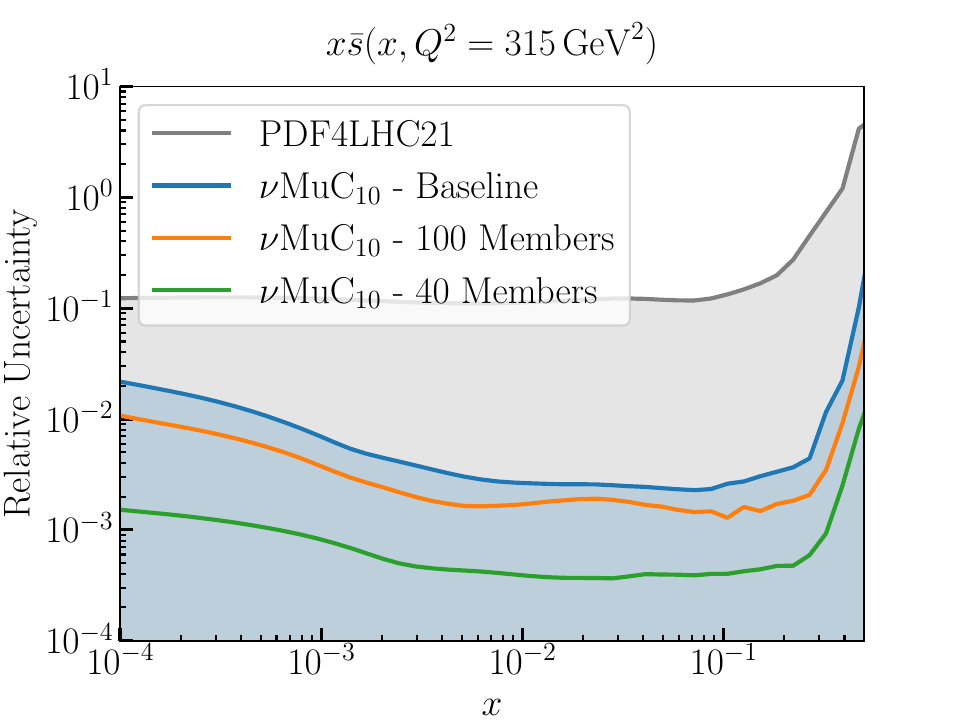} \\
    
    \includegraphics[width=0.47\textwidth]{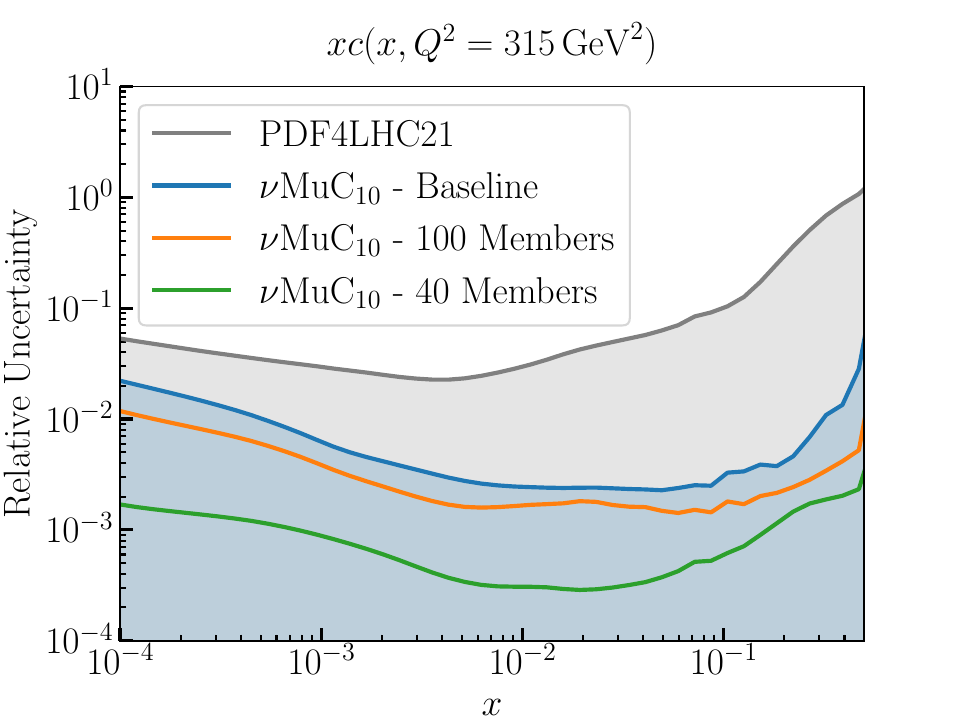} \hspace{-0.5cm}
    \includegraphics[width=0.47\textwidth]{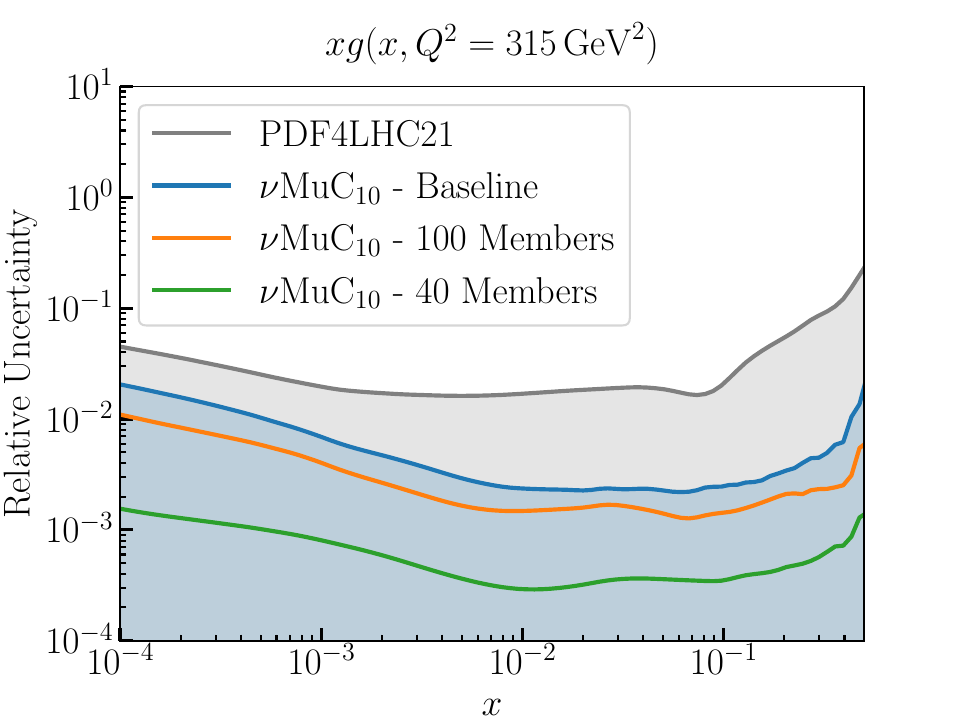}

    \caption{PDF uncertainties expected at the $\nuMuC$ experiment with 1500\,ab$^{-1}$ luminosity and at the 10\,TeV MuC, compared with current uncertainties from PDF4LHC21~\cite{PDF4LHCWorkingGroup:2022cjn}. 
    }
    \label{fig:pdf4lhc21_prior_post_1}
\end{figure}

Using this post-fit Hessian set, we compute the PDF uncertainties as a function of $x$ and for $Q^2=400\,{\rm{GeV}}^2$ using the baseline fit setup, 1500\,ab$^{-1}$ luminosity and 10\,TeV MuC neutrinos. The results are show in Figure~\ref{fig:pdf4lhc21_prior_post_1} in comparison with present-day uncertainties from PDF4LHC21~\cite{PDF4LHCWorkingGroup:2022cjn}. The uncertainties are below the percent level for all the partons, with an improvement of about one order of magnitude on the currently best-known PDFs and a much stronger improvement for the others. Note that although the gluon PDF is not directly probed in our analysis that is based on leading-order predictions, its determination also improves thanks to PDF sum rules, once quark and antiquark flavours are accurately measured. 

The results in the figure include profiling over the CKM matrix elements and the Weinberg angle, but we checked that this has a minor impact and very similar PDF uncertainties are obtained if the parameters are instead set to the PDG central value. The $\nuMuC$ data can disentangle and simultaneously determine the fundamental theory parameters and the PDFs.

The figure also displays the dependence on the number of members in the PDF Hessian set. With 40 members---i.e., with the same number of members of the default PDF4LHC21\_40 set---the attainable precision is considerably overestimated, while the results with 100 members are quite close to the baseline results obtained with 300 members. This saturation means that the 300 members set offers a flexible enough PDF parametrization, while the parametrization with 40 members introduces unphysical correlations among the different PDF functions, which are exploited in the fit to produce overly optimistic results. While this suggests that our baseline results are robust, we stress that ours is just an estimate and not an accurate sensitivity projection based on a proper PDF fit. Our methodology is `biased', in that our Hessian set parametrization builds on a Monte Carlo replica set that is constructed to model current data and their uncertainties. A more robust approach is to fit the PDFs from scratch taking into account the much more precise future $\nuMuC$ data in the selection of the PDF modelling (for instance, by neural networks). This is currently in progress~\cite{inprogPDF}. A final disclaimer, which applies to all our results, is that ours are estimates of the sensitivity floor from irreducible parametric uncertainties. The impact of theory uncertainties and of experimental systematics is still to be quantified.

\subsection{Sensitivity of a first MuC stage at \texorpdfstring{3\,TeV}{c}}
\label{sec:3TeVres}

\begin{figure}[t]
    \centering
    \includegraphics[width=0.46\textwidth]{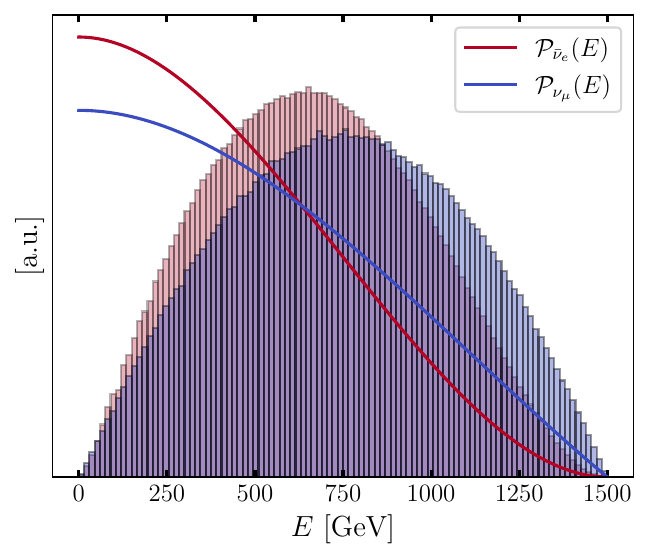}
    \hfill
    \includegraphics[width=0.5\textwidth]{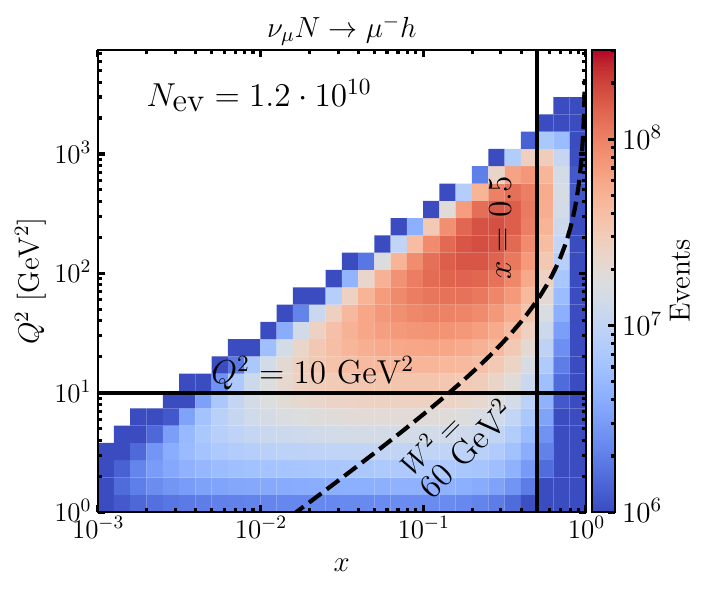}
    \caption{{\textbf{Left:}} Energy distribution of the interacting neutrinos, compared with the neutrino beam energy spectrum. {\textbf{Right}}: Number of events distribution (with high luminosity~(\ref{eq:HighL})) in $x$-$Q^2$ bins.}
    \label{fig:illustrative-3TeV}
\end{figure}

A 10\,TeV muon collider could be built in stages, by exploiting the muon cooling complex and the first part of the acceleration complex for a lower-energy collider in a dedicated ring. A 3\,TeV centre of mass energy option is considered in the IMCC design study, with the parameters reported in Table~1.1.1.~of Ref.~\cite{InternationalMuonCollider:2025sys}. The estimate of the luminosity of a $\nuMuC$ experiment performed at a 3\,TeV muon collider stage proceeds like in Section~\ref{ssec:ExS}, with the only difference that the reduced collider ring circumference---from 10\,km to $4.5$\,km---increases the rate $R_\nu$ for the decay of the neutrinos in the 10 meters long straight section of interest. This produces a higher integrated luminosity than in the 10\,TeV estimate, namely
\begin{equation}
    \int\mathcal{L}_\textrm{low} = 33\,\textrm{ab}^{-1}\,,\qquad
    \int\mathcal{L}_\textrm{high} = 3300\,\textrm{ab}^{-1}\,,
\end{equation}
depending on assumptions on the structure and the mass of the target. The luminosity is higher than at 10\,TeV (for the same assumed operation time of 5~years) but the number of collected DIS events is smaller because the cross section is smaller for less energetic neutrinos. The total event yields and the kinematic distributions are shown in Figure~\ref{fig:illustrative-3TeV} for the 3\,TeV $\nuMuC$ experiment (with high $3300\,\textrm{ab}^{-1}$ luminosity) and they can be compared with the 10\,TeV case in Figure~\ref{fig:illustrative}. 

While they do not have a direct impact on our results, two shortcomings of the reduced energy of the muons at the 3\,TeV collider are worth mentioning. First, the beam energy reduction by a factor of around 3 raises the intrinsic angular spread of the neutrino beam---from the muon decay process---by the same factor, such that $\langle\theta\rangle\sim0.1$\,mrad at the 3\,TeV collider. Correspondingly, the optimal placement of a 10\,cm radius cylindrical target would be about 600\,m from the IP and not 2\,km as in the 10\,TeV case. Incorporating the experiment in the design could thus be harder than at 10\,TeV. Additionally, the reduction of the muon energy reduces the energy of the neutrino interactions and it lowers the $Q^2$ distribution of the DIS events. The $Q^2$ distribution is now peaked---see Figure~\ref{fig:illustrative-3TeV}---at about 100\,GeV$^2$, which is still within the domain of perturbative QCD, but lower than at 10\,TeV.

The sensitivity projections on the CKM parameters at the $\nuMuC$ experiment exploiting the 3\,TeV MuC neutrinos are reported in Table~\ref{tab:CKMPDG}. They are obtained with a fit that employs the same observables discussed in Section~\ref{sec:CKMd} and the same modeling of the parametric uncertainties as in Section~\ref{subsec:Parametric_Uncertainties}. The binning in $x$ is the same---30 logarithmically spaced bins for $x \in [10^{-3},1]$, with the last 3 bins discarded---while $Q^2$ ranges from 10~GeV$^2$ to 3000~GeV$^2$ in 24 logarithmically spaced bins. The neutrino energy $E \in [0,1.5]$~GeV is divided into 5 linearly spaced bins.

The CKM sensitivity at 3\,TeV is only slightly worse than the 10\,TeV sensitivity. A significant progress is possible on the CKM parameters determination, in comparison with current knowledge, also at a 3\,TeV muon collider stage. 

\section{Conclusions and outlook}
\label{sec:conc}

We have established---see Sections~\ref{subsec:FitRes} and~\ref{sec:3TeVres}---the potential of a neutrino experiment ($\nuMuC$) parasitic to high-energy muon colliders for a radical improvement in the knowledge of the CKM quark mixing parameters well above current standards and with a novel methodology that is independent from previous low-energy determinations. The result is based on a comprehensive global analysis of deep inelastic scattering processes, modelled by perturbative QCD supplemented with the phenomenological parametrization of the PDFs and of the FFs that are determined simultaneously with the CKM parameters in the fit. The analysis is based on many ratios of binned cross sections (see Table~\ref{tab:Oblist}) constructed out of CC and NC measurements in the inclusive DIS and in bottom, charm and strange tagged SIDIS final states. Measurements in all these different final states are needed for a precise global determination of the CKM parameters.

We also quantified (Section~\ref{sec:PDFdet}) the excellent perspectives of the $\nuMuC$ experiment for the determination of the PDFs way beyond current precision. This result emerges as a byproduct of our fit by studying the posterior constraints on the PDF nuisance parameters. A similar advance is expected in the determination of the FFs, but it would require observables that are binned also in the SIDIS momentum fraction $z$, which is instead treated inclusively in our analysis. 

An important caveat is that our PDF determination employs an Hessian set modelling that is ultimately based on current PDF data and not designed to accommodate the much more precise future $\nuMuC$ data. We used a custom extended set with 300 members and we studied the dependence on the number of members in order to quantify the effect on the results of this bias in the PDFs parametrization. We found that the impact is moderate on the PDF fit and very small on the CKM parameters precision. Nevertheless, we consider that performing a full-fledged PDF fit is needed for a fully conclusive assessment. Along similar lines, the modelling of the FFs should be also refined and adapted to the high expected precision of the $\nuMuC$ data. 

Our results assume exact isospin symmetry and ignore nuclear effects. Their inclusion will definitely enlarge the uncertainties or even prevent the simultaneous determination of the PDFs for protons and for neutrons. Our estimate effectively corresponds to the expected precision on the determination of the PDFs of the entire nucleus (e.g., tungsten or silicon) that composes the target, and not of the individual constituents nucleons. Such nuclear PDFs control the physical cross sections and they will be determined together with the CKM parameters in the fit. Nuclear and isospin breaking effects can limit the precision on the extraction of the proton and neutron PDFs, but they will have no impact on the CKM parameters precision.

Ours is the first exploration\,\footnote{A preliminary version of our results was presented in Ref.~\cite{InternationalMuonCollider:2025sys}.}\,of the opportunities offered by deep inelastic scattering measurements with the neutrinos produced by a muon collider of very high energy. These opportunities are a significant addition to the already established muon collider physics potential exploiting muon collisions. While the best results are obtained with 10\,TeV muon collider neutrinos, a possible first muon collider stage with 3\,TeV energy in the centre of mass is also capable (see Table~\ref{tab:CKMPDG}) of a strong advance in the CKM parameters knowledge. Our results motivate future detailed studies that connect with the ongoing neutrino far-forward physics program at the LHC and HL-LHC offering an appealing long-term perspective of improvement for this active field of research.

\subsubsection*{Future work}

Ours is an estimate of the sensitivity floor set by the lack of knowledge of the PDFs and of the FFs, which need to be determined simultaneously with the CKM parameters from the same data. Assessing the actual perspectives of attaining this floor requires work on phenomenology, experimental physics and QCD theory. The advances needed in these three areas are discussed below in turn.

The aforementioned bias in the PDF parametrization should be eliminated by deploying a proper PDF fit methodology (such as for instance NNPDF) that adapts the PDF modelling to the very precise future $\nuMuC$ data. The possible impact of a more refined FF modelling should be also investigated. A further analysis improvement---which requires some advance in the standard PDF fit workflow---is the inclusion of smearing effects associated with the finite resolution in the measurement of the energy and the momentum of the final state hadrons and leptons, setting targets for the detector design. Finally, the analysis can be straightforwardly extended to the NLO in QCD, enabling an assessment of the impact of gluon production on the CKM parameters sensitivity. Work along all these lines is currently in progress~\cite{inprogPDF}.

Experimental physics work is needed on the experiment design. A robust starting point will be provided~\cite{Calzolari2025Neutrino,Calzolari2025MDI} by the detailed simulation of the neutrino flux based on the muon collider interaction region design. Preliminary  targets for the experiment that emerge from our study are a luminosity of about 1500\,ab$^{-1}$ enabling (see Figure~\ref{fig:luminosity-gainy}) a very strong advance in the CKM parameters determination, as well as excellent particle identification capabilities to tag the flavour of bottom, charmed and strange hadrons in order to select the corresponding SIDIS final states. More refined specification requests such as resolutions (and the calibration uncertainties) will emerge from the analysis refinements previously described.

The statistical precision on some of our observables is as small as $10^{-4}$. The observables are ratios of binned cross sections, in which pure QCD radiative corrections cancel out but only up to finite bin width effects. Furthermore, quark mass effects and QED radiative effects are definitely important at this high level of precision and even non-perturbative power corrections could be relevant. Theoretical predictions for the $\nuMuC$ experiment pose at once significant challenges and great opportunities for QCD theoretical calculations.

\section*{Acknowledgments}
We thank Felix Ringer for providing access to the $D^*$ fragmentation function grids, and Terry Generet and Rene Poncelet for discussions on their $B$-hadron fragmentation functions. We are indebted with Daniele Calzolari, Stefano Forte, Emanuele Nocera and Juan Rojo for the useful discussions. DM acknowledges support by the Italian Ministry of University and Research (MUR) via the PRIN project n.~20224JR28W. This work is part of the doctoral thesis of FM, within the framework of the Doctoral Program in Physics of the Autonomous University of Barcelona. The
work of FM is supported by the Joan Oró predoctoral grant program of the Department of Research and Universities of the Government of Catalonia and co-funded by the European Social Fund Plus (Project reference: 2024 FI-1 00715). This work is also part of the R\&D\&i project PID2023-146686NB-C31, funded by MICIU/AEI/10.13039/501100011033/ and by ERDF/EU. It is also supported by the Departament de Recerca i Universitats from Generalitat de Catalunya to the Grup de Recerca 00649 (Codi: 2021 SGR 00649). IFAE is partially funded by the CERCA program of the Generalitat de Catalunya. We also acknowledge financial support from the Spanish Ministry of Science and Innovation (MICINN) through the Spanish State Research Agency, under Severo Ochoa Centres of Excellence Programme 2025-2029 (CEX2024001442-S).

\appendix

\section{DIS and SIDIS kinematics and cross sections}
\label{app:DISAPP}

\subsection*{Kinematics}

The DIS and SIDIS processes of interest can be described by the following Lorentz invariant variables:
\begin{equation}\label{eq:kinv}
    E=\frac{p\cdot k}{M}\,,\qquad
    x=\frac{k\cdot k^\prime}{p\cdot(k-k^\prime)}\,,\qquad y= \frac{p\cdot (k-k^\prime)}{p\cdot k}\,,\qquad 
    z=\frac{p\cdot p_{{F}}}{p\cdot (k-k^\prime)}\,,
\end{equation}
where $k$ is the 4-momentum of the incoming neutrino, $k^\prime$ is the momentum of the outgoing charged lepton (considered massless) or neutrino, and $p=(M,\vec{0})$ is the nucleon 4-momentum. In the equation, $p_{{F}}$ is the 4-momentum of the $s$-, $c$- or $b$-flavoured hadron that is present in the SIDIS final state. The total 4-momentum of the produced hadrons will be denoted as $p_h=k-k^\prime+p$. The variable $E$ is the energy of the incoming neutrino in the lab frame and it is bounded in the interval $[0,\bar{E}]$, with $\bar{E}\simeq\ $5\,TeV ($1.5$\,TeV) for the 10\,TeV (3\,TeV) MuC. The variables $x$, $y$ and $z$ are the regular DIS variables and are bounded from~0 to~1.

The $(E,x,y,z)$ variables are in one-to-one correspondence with the following lab-frame quantities: the energy $E_\ell$ of the outgoing charged lepton or neutrino and its polar emission angle, $\theta_\ell$; the total energy $E_h$ of the hadrons; the energy $E_{F}$ of the heavy-flavoured hadron that is tagged in the SIDIS processes. These variables can be expressed as 
\begin{eqnarray}
&
E_\ell=(1-y)E\,,\qquad
&\sin^2\left({\theta_\ell}/{2}\right) =
\frac{xy\,M}{2E(1-y)}\,,\\
&E_h=M+y\,E\,,\qquad
&E_{F}=z\,y\,E\,.
\end{eqnarray}
Notice that by the fact that $\sin^2\left({\theta_\ell}/{2}\right)\leq1$, we get a constraint on $x$
\begin{equation}\label{eq:ct1}
    x\leq\frac{2E}{M}\frac{1-y}{y}\,.
\end{equation}
The variable $y$ is convenient for Monte Carlo integration, but it is often traded for 
\begin{equation}
Q^2=-(k-k^\prime)^2=2 k\cdot k^\prime=2MxyE\,,
\end{equation}
which has a more clear physical interpretation as the scale of the hard partonic scattering. Another important quantity is the `inelasticity' 
\begin{equation}\label{eq:Wq}
W^2=M^2+Q^2\frac{1-x}{x}=p_h^2=M^2+2ME\,y (1-x)\,,
\end{equation}
which is the total invariant mass of the hadrons in the final state. The deeply inelastic scattering regime is when the inelasticity is much larger than the nucleon mass squared $M^2$. If additionally $Q^2$ is much large than the hadronic scale, the DIS and SIDIS scattering cross sections can be predicted using perturbative QCD.

The variables $x$, $Q^2$ and $z$ can be expressed in terms of lab-frame measurable quantities as in Eq.~(\ref{eq:kinvar0}). As they only depend on the hadrons' momenta, they can be measured both in the CC and in the NC scattering events. The neutrino energy $E=E_\ell+E_h-M$ is instead only accessible in the CC processes by measuring the energy of the charged lepton. 

\subsection*{Cross sections}

The cross sections at the tree-level order and in the 5-flavour scheme can be obtained from the following building blocks
\begin{equation}\label{eq:buildingblock}
\begin{split}
& 
d\sigma(\nu_- q_-\to \ell_- q^\prime_- )=
d\sigma(\bar\nu_+ {\bar{q}}_+\to \bar\ell_+ {\bar{q}}^\prime_+)=
\frac{4G^2MEx}{\pi(1+Q^2/m_V^2)^2}\cdot{dxdy}\,,\\
& 
d\sigma(\bar\nu_+ q_-\to \bar\ell_+ q^\prime_- )=
d\sigma(\nu_- {\bar{q}}_+\to \ell_- {\bar{q}}^\prime_+)=
\frac{4G^2MEx(1-y)^2}{\pi(1+Q^2/m_V^2)^2}\cdot{dxdy}\,,
\end{split}
\end{equation}
that describe the polarised cross sections (the particle helicity $\pm 1/2$ is denoted as $\pm$) for massless quarks $q$ and $q^\prime$ scattering with neutrinos and producing a charged lepton or neutrino $\ell$. The formulas account for the exchange a generic vector of mass $m_V$ with vector-like coupling $g/\sqrt{2}$ to the leptons and to the quarks and the dimensionful coupling $G$ is defined as $G/\sqrt{2}=g^2/8m_V^2$ in analogy with the Fermi constant. In the CC and NC processes, $m_V=m_W$ and $m_V=m_Z$, respectively.

Multiplying by the PDFs we readily obtain the triply-differential cross sections, defined as in Eq.~(\ref{eq:Sigma}), for the CC processes:
\begin{eqnarray}
&&
\begin{aligned}
\frac{d\sigma({\nu_\mu N\to\mu^-h})}{dE dx dy} 
=   {\mathcal{P}}_{\nu_\mu} (E) \frac{2 G_F^2 M E x}{\pi(1+Q^2/m_W^2)^2} ~
&\bigg[\sum\limits_{\vphantom{\bar{d}}\textrm{I}=d,s,b}\sum\limits_{\vphantom{\bar{d}}\textrm{F}=u,c}
    |V_{\textrm{F}\textrm{I}}|^2 \textrm{I}(x,Q^2)+
\label{eq:CCnu}\\ 
    & \hspace{-40pt} +\sum\limits_{\vphantom{\bar{d}}\textrm{I}=\bar{u},\bar{c}}\sum\limits_{\vphantom{\bar{d}}\textrm{F}=\bar{d},\bar{s},\bar{b}}
    |V_{\textrm{I}\textrm{F}}|^2
    (1-y)^2
    {{\textrm{I}}}(x,Q^2)
     \bigg],
\end{aligned}\\
&&\begin{aligned}
\frac{d\sigma({\bar\nu_e N\to e^+ h})}{dE dx dy} 
=   {\mathcal{P}}_{\bar\nu_e} (E) \frac{2 G_F^2 M E x}{\pi(1+Q^2/m_W^2)^2} ~ & 
\bigg[\sum\limits_{\vphantom{\bar{d}}\textrm{I}=\bar{d},\bar{s},\bar{b}}
    \sum\limits_{\vphantom{\bar{d}}\textrm{F}=\bar{u},\bar{c}}
    |V_{\textrm{F}\textrm{I}}|^2 
    {{\textrm{I}}}(x,Q^2)+
\label{eq:CCnubar}\\ 
    & \hspace{-37pt}    +\sum\limits_{\vphantom{\bar{d}}\textrm{I}=u,c}
    \sum\limits_{\vphantom{\bar{d}}\textrm{F}=d,s,b}
    |V_{\textrm{I}\textrm{F}}|^2 (1-y)^2\textrm{I}(x,Q^2)
 \bigg],
\end{aligned}
\end{eqnarray}
where `$V$' denotes the CKM matrix. The two sums runs over the initial and final state quarks or anti-quarks that are involved in the reaction. The top quark is not kinematically accessible and hence it does not appear in the final state. The PDFs of the initial state quarks and anti-quarks are represented by ${\textrm{I}}(x,Q^2)$.

The NC cross section is the sum of the contributions from $\nu_\mu$ and $\bar\nu_e$ scattering. Taking into account the couplings $g_f = T^3_L(f) - Q_f \sin^2 \theta_{\textsc{w}}$ of the $Z$ boson, the cross sections read
\begin{eqnarray}
&&
\begin{aligned}
\hspace{-30pt}\frac{d\sigma({\nu_\mu N\to\nu_\mu h})}{dE dx dy} 
=   {\mathcal{P}}_{\nu_\mu} (E) \frac{2 G_F^2 M E x}{\pi(1+Q^2/m_Z^2)^2} ~ &
\bigg[\sum\limits_{\vphantom{\bar{d}}\textrm{I}=q} 
    \left( g_{q_L}^2 + g_{q_R}^2(1-y)^2 \right)
    \textrm{I}(x,Q^2)\,+
\label{eq:NCnu}\\ 
    & \hspace{-0pt}+     \sum\limits_{\vphantom{\bar{d}}\textrm{I}=\bar{q}} 
    \left( g_{q_R}^2 + g_{q_L}^2(1-y)^2 \right)
    \textrm{I}(x,Q^2)\bigg],
\end{aligned}\\
&&\begin{aligned}
\hspace{-30pt}\frac{d\sigma({\bar\nu_e N\to\bar\nu_e h})}{dE dx dy} 
=   {\mathcal{P}}_{\bar\nu_e} (E) \frac{2 G_F^2 M E x}{\pi(1+Q^2/m_Z^2)^2} ~ &
\bigg[\sum\limits_{\vphantom{\bar{d}}\textrm{I}=q} \left( g_{q_R}^2 + g_{q_L}^2(1-y)^2 \right)
    \textrm{I}(x,Q^2)\,+
\label{eq:NCnubar}\\ 
    & \hspace{-0pt}\times +    \sum\limits_{\vphantom{\bar{d}}\textrm{I}=\bar{q}} 
    \left( g_{q_L}^2 + g_{q_R}^2(1-y)^2 \right)
    \textrm{I}(x,Q^2)\bigg],
\end{aligned}
\end{eqnarray}
where the sum over the initial state parton runs separately over quarks $q=u,c,d,s,b$ and anti-quarks $\bar{q}=\bar{u},\bar{c},\bar{d},\bar{s},\bar{b}$. The final state is the same as the initial state since the $Z$ interactions are diagonal. 

The SIDIS cross sections are readily obtained by including the FFs in the corresponding DIS cross section formulas. The two CC cross sections for a generic tagged hadron $F$ are
\begin{eqnarray}
\raisetag{50\baselineskip}
&&
\begin{aligned}
\hspace{-20pt}\frac{d\sigma({\nu_\mu N\to\mu^-F})}{dE dx dy dz} 
=   {\mathcal{P}}_{\nu_\mu} (E) \frac{2 G_F^2 M E x}{\pi(1+Q^2/m_W^2)^2} ~ &\bigg[\sum\limits_{\vphantom{\bar{d}}\textrm{I}=d,s,b}\sum\limits_{\vphantom{\bar{d}}\textrm{F}=u,c}
    |V_{\textrm{F}\textrm{I}}|^2 \textrm{I}(x,Q^2){\mathcal{F}}_{\textrm{F}}(z,Q^2)\,+
\label{eq:CCnuB}\\ 
    & \hspace{-50pt}+\sum\limits_{\vphantom{\bar{d}}\textrm{I}=\bar{u},\bar{c}}\sum\limits_{\vphantom{\bar{d}}\textrm{F}=\bar{d},\bar{s},\bar{b}}
    |V_{\textrm{I}\textrm{F}}|^2 
    (1-y)^2
    {{\textrm{I}}}(x,Q^2){\mathcal{F}}_{\textrm{F}}(z,Q^2)
     \bigg],
\end{aligned}\\
&&\begin{aligned}
\hspace{-20pt}\frac{d\sigma({\bar\nu_e N\to e^+ F})}{dE dx dy dz} 
=   {\mathcal{P}}_{\bar\nu_e} (E) \frac{2 G_F^2 M E x}{\pi(1+Q^2/m_W^2)^2} ~ &
\bigg[\sum\limits_{\vphantom{\bar{d}}\textrm{I}=\bar{d},\bar{s},\bar{b}}
    \sum\limits_{\vphantom{\bar{d}}\textrm{F}=\bar{u},\bar{c}}
    |V_{\textrm{F}\textrm{I}}|^2 
    {{\textrm{I}}}(x,Q^2){\mathcal{F}}_{\textrm{F}}(z,Q^2)\,+
\label{eq:CCnubarB}\\ 
    & \hspace{-50pt}     +\sum\limits_{\vphantom{\bar{d}}\textrm{I}=u,c}
    \sum\limits_{\vphantom{\bar{d}}\textrm{F}=d,s,b}
    |V_{\textrm{I}\textrm{F}}|^2 
    (1-y)^2\textrm{I}(x,Q^2){\mathcal{F}}_{\textrm{F}}(z,Q^2)
     \bigg].
\end{aligned}
\end{eqnarray}
Similarly, the NC SIDIS cross section is the sum of the $\nu_\mu$ and $\bar\nu_e$ contributions
\begin{eqnarray}
&&
\begin{aligned}
\hspace{-20pt}\frac{d\sigma({\nu_\mu N\to\nu_\mu F})}{dE dx dy dz} 
=   {\mathcal{P}}_{\nu_\mu} (E) \frac{2 G_F^2 M E x}{\pi(1+Q^2/m_Z^2)^2} &
\bigg[\sum\limits_{\vphantom{\bar{d}}\textrm{I}=q} 
    \left( g_{q_L}^2 + g_{q_R}^2(1-y)^2 \right)
    \textrm{I}(x,Q^2){\mathcal{F}}_{\textrm{I}}(z,Q^2)
\label{eq:NCnuB}\\ 
    & \hspace{-30pt}+    \sum\limits_{\vphantom{\bar{d}}\textrm{I}=\bar{q}} 
    \left( g_{q_R}^2 + g_{q_L}^2(1-y)^2 \right)
    \textrm{I}(x,Q^2){\mathcal{F}}_{\textrm{I}}(z,Q^2)\bigg],
\end{aligned}\\
&&\begin{aligned}
\hspace{-20pt}\frac{d\sigma({\bar\nu_e N\to\bar\nu_e F})}{dE dx dy dz} 
=   {\mathcal{P}}_{\bar\nu_e} (E) \frac{2 G_F^2 M E x}{\pi(1+Q^2/m_Z^2)^2} &
\bigg[\sum\limits_{\vphantom{\bar{d}}\textrm{I}=q} 
    \left( g_{q_R}^2 + g_{q_L}^2(1-y)^2 \right)
    \textrm{I}(x,Q^2){\mathcal{F}}_{\textrm{I}}(z,Q^2)
\label{eq:NCnubarB}\\ 
    & \hspace{-30pt}+    \sum\limits_{\vphantom{\bar{d}}\textrm{I}=\bar{q}} 
    \left( g_{q_L}^2 + g_{q_R}^2(1-y)^2 \right)
    \textrm{I}(x,Q^2){\mathcal{F}}_{\textrm{I}}(z,Q^2)\bigg].
\end{aligned}
\end{eqnarray}
In the equations, ${\mathcal{F}}_{\textrm{F}}(z,Q^2)$ denotes the appropriate FF---see Appendix~\ref{app:pdf_ff}---for the hadron flavour $F$ and the final state quark or anti-quark ${\textrm{F}}$ under examination. 

\section{Parton distribution functions and fragmentation functions}
\label{app:pdf_ff}

In this appendix we describe the Hessian sets that we employ to model parametric uncertainties from the imperfect knowledge of the PDFs and of the FFs. Particular attention is devoted to the PDFs, aiming at producing new and more expressive Hessian sets based on the PDF4LHC21 combination~\cite{PDF4LHCWorkingGroup:2022cjn}, beyond the public LHAPDF~\cite{Buckley:2014ana} releases (which currently provide a 40-member Hessian set and a 100-replica Monte Carlo set).

\subsection*{Parton distribution functions}

Among the most widely used modern global PDF sets are NNPDF4.0~\cite{NNPDF:2021njg}, MSHT20~\cite{Bailey:2020ooq}, and CT18~\cite{Hou:2019efy}. Although these state-of-the-art sets represent the current standard in the field, they are built upon different methodological assumptions and fitting strategies. Such differences, ranging from dataset selection to parametrisation choices and uncertainty treatments, can lead to non-negligible discrepancies in the extracted PDFs, which in turn propagate to phenomenological analyses.

\begin{figure}[t]
    \centering   
    \includegraphics[width=0.65\linewidth]{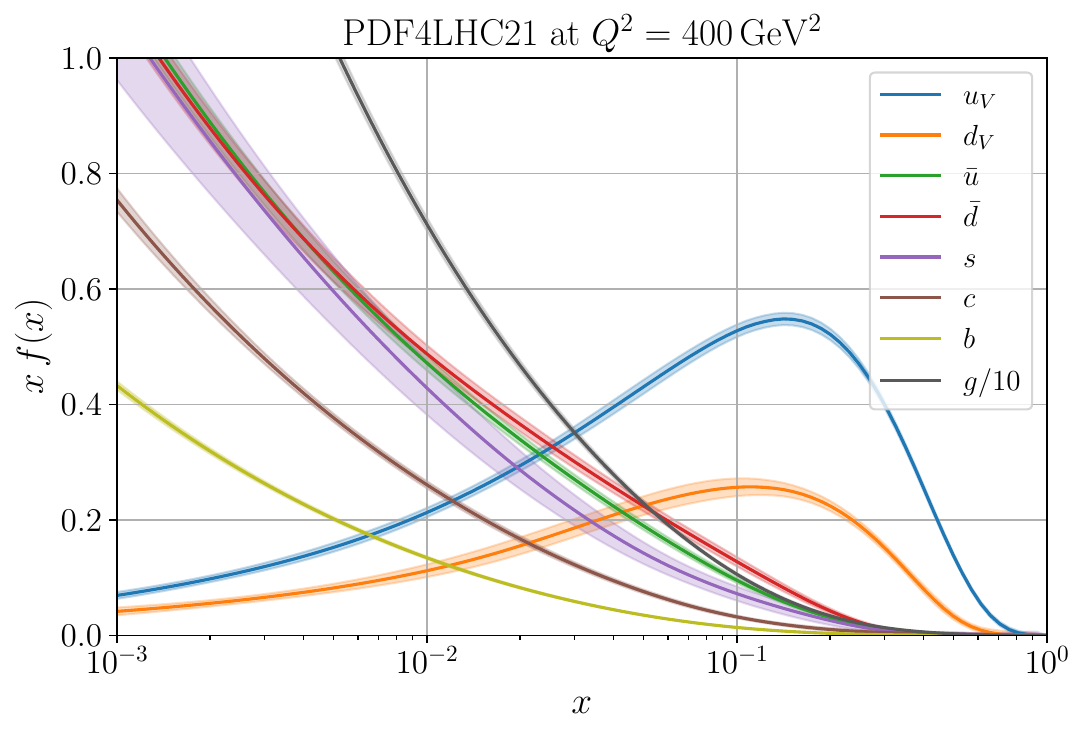}
    \caption{PDF4LHC21 for different channels at $Q^2 =400 \,  \rm{GeV}^2$.}
    \label{fig:pdf_summary}
\end{figure}

To account for these methodological differences and provide a more conservative, agnostic benchmark, the PDF4LHC Working Group has produced the PDF4LHC21 combination~\cite{PDF4LHCWorkingGroup:2022cjn}, which incorporates results from the three aforementioned collaborations (using NNPDF3.1~\cite{NNPDF:2017mvq} in place of NNPDF4.0). In this construction, the spread of the combined set reflects not only experimental and theoretical uncertainties, but also legitimate methodological differences, thereby offering a more conservative PDF set than any single global fit. For this reason, and in order to avoid bias towards a particular fitting philosophy, we adopt PDF4LHC21 in our analysis. Figure~\ref{fig:pdf_summary} provides an overview of the PDF4LHC21 PDFs at $Q^{2}=400~\mathrm{GeV}^{2}$, showing the enhancement of valence quarks at large~$x$ and the dominance of sea quarks and the gluon at small~$x$.

The PDF4LHC21 combination is constructed from the three global fits CT18, MSHT20, and NNPDF3.1. These fits are performed on a reduced common dataset including both DIS and hadronic data. This homogenisation of input data allows to isolate methodological differences between the groups while retaining sensitivity to key experimental constraints. Since NNPDF3.1 provides native Monte Carlo (MC) replicas while MSHT20 and CT18 are Hessian-based, the latter are converted to MC format using the symmetrised Watt--Thorne method~\cite{Watt:2012tq}. Each fit contributes 300 MC replicas for a total of 900. Finally, two compressed representations of the full 900 MC replicas set are produced: PDF4LHC21\_mc with 100 MC replicas and the Hessian set PDF4LHC21\_40, with 40 members plus the central value. The PDF4LHC21\_40 set is generated by the MC-to-Hessian conversion method META-PDF~\cite{Gao:2013bia}. A different conversion method, MC2Hessian~\cite{Carrazza:2015aoa,Carrazza:2016htc}, was also studied giving similar performances and it will be employed in what follows to produce new Hessian sets out of the 900 MC replicas PDF4LHC21 set.

The PDF central values and uncertainties can be represented through summary statistics of the PDF MC ensemble. The central value of the Hessian set is simply the mean of the MC replicas. The construction of the Hessian set members amounts to finding a multi-Gaussian representation of uncertainties in PDF space, requiring only the covariance matrix. In the MC2Hessian method---see the original references~\cite{Carrazza:2015aoa,Carrazza:2016htc} for a complete discussion---this is achieved as follows.

The PDF functions can be represented by their discretization on a narrow grid in $x$, namely they are represented by their values on a large number $N_x$ of grid points $x=x_i$, with $i=1,\dots,N_x$. The discretization is performed at fixed $Q^2=Q_0^2$ and the dependence on $Q^2$ is obtained by solving the Altarelli--Parisi evolution equations. In particular, we set $Q_0=30$\,GeV corresponding to the typical scale of the $\nuMuC$ DIS data. In this representation, the fluctuations around the central value of the PDFs across the MC replica set form a matrix $X$ called the sampling matrix, which is given by 
\begin{equation}
X_{l,\,{\textsc{m}}} \equiv {\hat{f}}^{({\textsc{m}})}_{\alpha}(x_i,Q^2_0) - f^{(0)}_{\alpha}(x_i,Q^2_0),
\end{equation}
where the index $\alpha=1,\dots,N_f$ runs over flavours and $i$ over the grid points. The two indices are flattened in the single index $l=N_x(\alpha-1)+i$, which runs from 1 to $N_x\cdot N_f$. The index ${\textsc{m}}=1,\dots,N_{\text{rep}}$ runs over the replicas of the MC set and the matrix $X$ has dimension $(N_x\cdot N_f)\times  N_{\text{rep}}$. The PDF uncertainties are provided by the $(N_x\cdot N_f)\times (N_x\cdot N_f)$ covariance matrix
\begin{equation}\label{eq:covm}
\text{cov} = \frac{1}{N_{\text{rep}}-1}\, XX^t.
\end{equation}

The MC2Hessian method aims at defining an Hessian set whose covariance is to a good approximation equal to Eq.~(\ref{eq:covm}). In order to construct the set, we perform the Singular Value Decomposition of the sampling matrix 
\begin{equation}
\label{eq:svd}
 X = USV^t,
\end{equation}
obtaining two orthogonal matrices $U$ and $V$ and the diagonal matrix $S$ of the singular values. We order the singular values, $s_k$, from largest to smallest. Notice that each column `$k$' of $U$ is the Eigenvector of the covariance matrix~(\ref{eq:covm}), with Eigenvalue $s_k^2$. There are in general `$N_{\text{rep}}$' non-vanishing Eigenvalues because this is the rank of the covariance matrix.

The members of the Hessian set are defined in terms the values they assume on the $x$ grid point and at fixed $Q=Q_0$ (the dependence on $Q$ is reconstructed by Altarelli--Parisi) by the following expression:
\begin{equation}\label{eq:HMS}
    f_\alpha^{(k)}(x_i,Q_0^2)=f_\alpha^{(0)}(x_i,Q_0^2)
    +s_k U_{N_x(\alpha-1)+i,\,k}\,,
\end{equation}
where $k=1,\ldots N_{\text{rep}}$ runs over the non-vanishing Eigenvalues $s_k$. It is possible to show that the uncertainties associated to a PDF Hessian set composed by all the $N_{\text{rep}}$ members defined in Eq.~(\ref{eq:HMS}) correspond exactly to the covariance matrix in Eq.~(\ref{eq:covm}). A good approximation of the covariance matrix can be obtained by truncating the matrix $S$ in Eq.~(\ref{eq:svd}) to its first $N$ largest values, or equivalently by considering only its first $N$ Eigenvectors with larger Eigenvalues. Correspondingly, we can construct Hessian sets with a variable number of members $N\leq N_{\text{rep}}$ using Eq.~(\ref{eq:HMS}) with $k$ ranging from 1 to $N$.

\subsection*{Fragmentation functions}

Fragmentation functions (FFs) are essential for predicting observables involving detected hadrons in the final state. Much like PDFs describe the partonic structure of incoming hadrons, FFs encode---in a simplified leading-order picture---the probability that an outgoing parton hadronises into a specific hadron type, which is ultimately measured in the detector.

The fragmentation function $\mathcal{F}_{p}(z, Q^2)$ describes the probability for a parton $p$ to produce a hadron $F$ carrying a fraction $z$ of the parton’s momentum at an energy scale $Q^2$. Like PDFs, FFs require non-perturbative input and must be constrained by experimental data. First, a functional form for the FF is assumed at an initial scale $Q_0^2$. Next, the DGLAP equations are used to evolve the FF to the relevant scale $Q^2$. Finally, the parameters of the initial scale function are determined by fitting predictions to experimental measurements, which allows backward inference on the FF’s non-perturbative structure.

For our analysis we need $B$-hadron, $D^*$- and $K$-meson FFs for the theoretical description of bottom, charm and strange tagged SIDIS, respectively. We employ the $B$-hadron FFs derived in Ref.~\cite{Czakon:2022pyz}, which achieve NNLO QCD accuracy with next-to-next-to-leading logarithmic (NNLL) soft gluon resummation. These FFs were determined through global fits to $e^+e^-$ collision data. The FF set is provided in LHAPDF format~\cite{Buckley:2014ana} using a MC replica representation with 100 members. We used this set to produce an Hessian set with 40 members using the MC2Hessian methodology explained above.

For the $D^*$-meson FFs, we use the NLO QCD-accurate set from Ref.~\cite{Anderle:2017cgl}, which was determined through global fits to $e^*e^-$, hadron-hadron, and in-jet fragmentation proton-proton scattering data. Our implementation utilises bare $(z, Q^2)$ grids provided directly by the authors, which we have interpolated using SciPy's general-purpose cubic spline routines~\cite{2020SciPy-NMeth}. The native FF set follows a Hessian error representation consisting of 1 central member and 8 pairs of asymmetric error Eigenvectors (corresponding to $+1\sigma$ and $-1\sigma$ variations for each direction). The $\pm1\sigma$ variations of our observables are approximately equal and opposite for 7 out of the 8 Eigenvectors, so that the corresponding nuisance parameters can be effectively treated at the linear order. The variation is instead not even approximately symmetric for one of the parameters, which is thus treated at the quadratic order (see Footnote~\ref{fn:NL}).

For $K^{\pm}$-tagged SIDIS we use the NLO MAPFF1.0 FFs of Ref.~\cite{AbdulKhalek:2021gbh}, which are determined from a simultaneous global fit to semi-inclusive $e^+e^-$ annihilation data and SIDIS. The FFs are parametrised using neural networks and the associated uncertainties are propagated via the MC replica method, with a baseline fit containing 200 replicas. As in the case of the $B$-hadron FFs, we then compress this baseline ensemble into a more compact Hessian representation with 160 members, which provides more computational efficiency while preserving the uncertainty information.

\vspace{1 cm}
\bibliographystyle{JHEP}
\bibliography{biblio}

\end{document}